\documentclass[twoside,11pt]{article}

\usepackage[preprint]{jmlr2e}

\usepackage[utf8]{inputenc}
\usepackage{amsmath}
\usepackage{xcolor}
\usepackage{booktabs, multirow, multicol}
\usepackage{subcaption}
\usepackage{enumitem}
\usepackage{listings}
\usepackage{lastpage}
\expandafter\def\expandafter\UrlBreaks\expandafter{\UrlBreaks\do\/\do\*\do\-\do\~\do\'\do\"\do\-}
\usepackage{placeins}

\jmlrheading{}{2026}{1-\pageref{LastPage}}{}{}{}{Burak Susam and Tingting Mu}

\ShortHeadings{Explainable Iterative Data Visualization Refinement}{Susam and Mu}
\firstpageno{1}

\begin{document}

\title{Explainable Iterative Data Visualization Refinement\\ via an LLM Agent}

\author{\name Burak Susam \email burak.susam@postgrad.manchester.ac.uk \\
       \name Tingting Mu \email tingting.mu@manchester.ac.uk \\
       \addr Department of Computer Science\\
       University of Manchester\\
       Manchester, M13 9PT, UK}

\maketitle

\begin{abstract}%
Exploratory analysis of high-dimensional data relies on embedding the data into a low-dimensional space, based on which a visualization plot is produced to uncover meaningful structures and to communicate geometric and distributional data characteristics.
Finding a suitable embedding algorithm configuration, particularly the hyperparameter setting, to produce a visualization plot that faithfully represents the underlying reality and encourages pattern discovery is labor-intensive and requires domain knowledge.
To ease this burden, we propose an agentic AI pipeline that integrates rigorous quantitative assessment with large language model (LLM) based approximation of qualitative human judgment and insight.
By treating visualization evaluation and hyperparameter optimization as a semantic task, our system generates a multi-faceted report that contextualizes quantitative metrics with descriptive summaries, and suggests actionable recommendations of algorithm configuration for refining data visualization.
Through an iterative refinement loop, the system rapidly produces a high-quality visualization plot in full automation.
\end{abstract}


\section{Introduction}
\label{sec:intro}

Exploratory data analysis of high-dimensional datasets relies heavily on embedding the data into low-dimensional spaces (typically 2D or 3D) to uncover latent structures. 
It is a common practice to use dimensionality reduction (DR) algorithms such as PCA, t-SNE \citep{vandermaaten08a} and UMAP \citep{mcinnes2020umap} to compute 2D or 3D embeddings, and then to visualize the embedded points on scatter plots so that the data become visually interpretable. 
The primary goal of this visualization process is not merely to partition data, but to achieve a holistic \textbf{understanding of diverse data characteristics}, e.g., identifying clusters, hierarchies,  anomalies, and continuous manifolds.
However, using DR algorithms for visualization is not a straightforward process as each algorithm involves its own hyperparameters  that govern the computed embeddings, and the outcomes are often highly sensitive to the hyperparameter configuration. 
Although default settings are typically provided with these algorithms, they are frequently sub-optimal choices and seldom produce  the most suitable visualization results in practical applications. 
In most cases, a careful tuning of hyperparameters is required for the considered algorithm, in order to obtain informative and high-quality visualization from the input raw data. 
Although grid search provides a simple and systematic search strategy, it  heavily relies on user preferences  for defining, e.g., search ranges and step sizes, and  is  computationally demanding  due to the enumeration of parameter combinations.
More sophisticated approaches based on Bayesian optimization, evolutionary optimization, and reinforcement learning    require  specialized  knowledge and dedicated design on algorithms.

In addition to the challenge of identifying suitable hyperparameter configuration, evaluating whether a given visualization is able to faithfully represent the  high-dimensional reality of the underlying geometric, topological and distributional characteristics of the high-dimensional data constitutes another challenge. 
A ``good looking'' visualization might not  be the most suitable representation of the underlying data, as the used DR algorithm can potentially create artifacts such as artificial clusters, false neighbors, and incorrect cluster hierarchies.
Current evaluation practices are bifurcated, separating quantitative metrics from qualitative visual assessment.

On one hand, quantitative metrics are employed for evaluation, including DR metrics (e.g., Stress, Trustworthiness) and, frequently, clustering metrics (e.g., Silhouette score) used as proxies for visual separability.
As objective scores, they suffer from significant limitations. First, they are often reductive, as a single scalar value cannot capture the nuance of complex topologies. 
Second, reliance on clustering-specific metrics implicitly assumes that the data should form discrete groups, penalizing valid non-convex structures, gradients, or continuous trajectories that are crucial for data understanding \citep{arbelaitz2013extensive}. 
Most critically, these numbers lack explanatory power.
For instance, a low Trustworthiness score can indicate  a problem, but   fails to explain \textit{where} or \textit{why} the distortion is occurring.

On the other hand, analysts rely on qualitative visual inspection of visualization plots. 
Humans are adept at recognizing complex Gestalt patterns, such as semantic relationships between neighboring groups or subtle outlier formations, which are however difficult to formalize algorithmically \citep{chari2023stop}. 
Additionally, visual inspection is subjective and susceptible to misinterpretation of DR artifacts, where the visual density or inter-cluster distances may not reflect the true high-dimensional relationships \citep{vandermaaten08a}.

Such disconnection leads to the following five core problems in data visualization and understanding:

\noindent
 \textbf{(i) Lack of Explanatory Power:} Quantitative metrics are not self-explanatory and fail to describe the nature of structural distortions.

 \noindent
 \textbf{(ii) Lack of Automated Hyperparameter Tuning:} Determining a satisfactory hyperparameter configuration requires human effort.  Even with a hyperparameter tuning algorithm, it still needs manual definition of some constraints and human effort in search algorithm design. 
 The process becomes computationally more expensive as the tuning becomes more fine-grained.

 \noindent
\textbf{(iii) Oversimplification of Structure:} Relying solely on clustering metrics ignores other vital structural forms, such as continuous manifolds or hierarchies.

\noindent
\textbf{(iv) Fragmented Evaluation:} Analysts must manually integrate insights from abstract metrics and visual patterns, lacking support from a unified framework.

\noindent
\textbf{(v) Ambiguity in Diagnosis:} It is difficult to discern whether a confused visualization stems from inherent data ambiguity or sub-optimal DR hyperparameters.

Our goal is to develop a visualization framework that is easy to use and efficient at producing an informative and high-quality visualization plot, addressing the above identified limitations.  
Specifically, we propose an iterative hyperparameter optimization and visualization refinement framework using an AI agent, taking advantage of the strong pattern recognition,  contextual reasoning, and broad domain knowledge encoded in a modern large language  model (LLM).
In our framework, referred to as AutoTuneVis, we prompt an LLM agent with a carefully designed context containing mathematical algorithms for computing metrics,   hierarchy diagrams, embedding coordinates and visualization plot, and ask the agent to tune hyperparameters of a given DR algorithm supported by appropriate rationale and reasoning process. 
The resulting visualization generated with the LLM-suggested hyperparameter setting should keep domain-relevant aspects while avoiding algorithm-related artifacts. 
We emphasize  system explainability  by  providing rationale for every change, particularly the reasoning and references to related quantitative metrics, hierarchy changes and visual effects for every parameter sweep.

To validate the efficacy of AutoTuneVis, we evaluate it on complex, high-dimensional single-cell RNA sequencing datasets. 
Experimental results show that the proposed agentic pipeline is able to rapidly navigate the hyperparameter search spaces of diverse dimensionality reduction algorithms, including t-SNE, UMAP, and PaCMAP, finding suitable hyperparameters typically within a few iterations. 
The results also show   that the LLM agent can successfully identify and resolve detailed visual artifacts such as artificial fragmentation and class dominance that rigid objective functions often overlook. 
AutoTuneVis effectively balances local neighborhood preservation with global geometric fidelity, generating biologically faithful, interpretable visualization plots in full automation while providing clear, actionable rationales for every refinement step.

\section{Related Work}
\subsection{DR and Visualization Evaluation}
While non-linear DR algorithms, notably t-SNE \citep{vandermaaten08a}, UMAP \citep{mcinnes2020umap} and PaCMAP \citep{wang2021understanding}, are powerful tools to analyze  high-dimensional data in terms of global topology, local geometric structure, and multi-scale cluster arrangement, etc., they are  sensitive to their hyperparameter configurations.
A suboptimal hyperparameter setting can typically lead  to visual artifacts, such as artificial cluster fragments or false  manifolds \citep{wattenberg2016how}. 
The low-dimensional embeddings computed  by these algorithms are often evaluated using structural preservation metrics such as Trustworthiness, Continuity and Kruskal's Stress \citep{venna2006local}. 
Although these metrics are capable of indicating distortion that occurs during the DR process, they   fail  to diagnose  specific visual artifacts.

\subsection{Unsupervised Hyperparameter Optimization}

Automating the selection of DR hyperparameters is a challenging problem due to  the highly subjective nature of visual interpretation, particularly in the absence of ground truth  to establish an objective standard to inform the evaluation.
Conventional hyperparameter optimization techniques, such as grid search and random search \citep{bergstra12a} or Bayesian optimization \citep{snoek2012practical}, require a strict, scalar objective function that can be either computed or measured. 
In the context of data visualization, a common practice is to  combine competing quality metrics, e.g., maximizing Trustworthiness while minimizing Stress. 
However, such rigid formulation can fail  to capture the nuanced, multi-faceted requirements of exploratory data analysis, 
where analysts might prioritize the faithful representation of specific cluster relationships over global geometric fidelity. 
In general, hyperparameter optimization methods grounded purely in numerical criteria are unlikely to replicate the qualitative and semantic reasoning that an expert human analyst applies during iterative refinement.
\subsection{LLMs as Autonomous Agents}
The advent of LLMs has catalyzed a shift from traditional Automated Machine Learning (AutoML) to agentic AI systems capable of complex reasoning, tool use, and iterative planning \citep{wang2023survey}.
In parallel, there are efforts to use agentic frameworks for data analysis and visualization purposes, employing frameworks to implement code and automate visualization \citep{yang2024matplotagent} and making visualization process prompt-driven \citep{chen2025coda}. However, hyperparameter optimization was not a primary challenge tackled by those studies. 
\citet{liu2025agenthpo} developed AgentHPO (Agent-based Hyperparameter Optimization) to provide a generic hyperparameter tuning framework.
It has been tested with result-driven machine learning tasks, but not applied to the data visualization domain, which does not have ground-truth outputs to measure accuracy directly.
\section{What and How to Inspect}

When looking into a scatter plot of 2D embeddings generated from high-dimensional data by a DR algorithm, one expects to establish a comprehensive   understanding of the data, by identifying and analyzing hierarchies of clusters,  cluster formation and separation,  data density variation and potential anomalies, as well as local neighborhood and manifold properties.
In addition to visual inspection facilitated by color, point size and line/curve based highlights, a visualization should  be accompanied by data analysis tools to formalize such patterns of interest.
Specifically, we can compute a cluster hierarchy   by applying a hierarchical clustering algorithm, e.g., UPGMA \citep{sokal1958}.
In addition to highlighting the clusters in color, the result can be separately visualized as a dendrogram (tree diagram), supplementing the scatter plot.
With respect to cluster arrangement,  quantitative metrics that assess cluster quality in terms of  intra-cluster tightness and inter-cluster separation, e.g., Silhouette scores \citep{rousseeuw1987}, can be presented.
To track data density variation and flag potential outliers, the local outlier factor (LOF) algorithm \citep{breunig2000lof} can be used, which compares the local data density of a point to the densities of its neighbors  and flags points with lower densities as outliers.
To examine and quantify characteristics of local neighborhood and manifold structure, we can calculate statistics that measure the agreement between the ground-truth neighbors indicated by the original high-dimensional data points and the observed neighbors based on the 2D embeddings, and assess consistency  between pairwise distances in the original and 2D spaces.
Examples of such statistics include  trustworthiness, continuity  and stress  scores \citep{venna2006local}, which we will further explain in the next section.

In our agentic visualization pipeline design, we  provide to the LLM agent a comprehensive collection of results produced by the aforementioned data analysis tools, to supplement the visualization plot and 2D embedding coordinates, providing a richer context for the agent to reason.

\section{Visualization Reliability}

In addition to visually interpreting and reasoning about a visualization plot, it is essential to understand the reliability of the information communicated by the 2D embeddings used to generate the visualization.
It is well known that a substantial dimensionality reduction  into two (or three) dimensions inevitably introduces information distortion when representing complex high-dimensional data patterns.
To evaluate the reliability of the 2D embeddings,  we consider metrics that examine both \textbf{global} and \textbf{local} structure preservation \citep{spearman1904,venna2006local}.  
Specifically, the global  metrics   assess the preservation of pairwise relationships across the entire dataset, while the  local metrics  assess the fidelity of neighborhood structures around each point. 
As shown in our later pipeline design, these metrics are used to encourage the LLM to produce more reliable visualization.

Our pipeline employs three global reliability  metrics, including the \textbf{Spearman Correlation Score} ($\rho_s$)  that computes the Spearman's rank correlation coefficient for distance ranks,  the \textbf{Stress Score} ($S$) that quantifies  distance distortion from the high-dimensional space to the low-dimensional space,  and the \textbf{Mean Distance Ratio} ($R$) that focuses on assessing the scaling change of pairwise distance from the high-dimensional space to low-dimensional space.
Specifically, a higher  $\rho_s $ (e.g.,  closer to 1)  indicates that the relative ordering of the pairwise distances is  overall well preserved for all pairs of data  points.
A lower $S$ indicates a better preservation of the original data geometry.
A value of $R$ close to 1 indicates a low distortion of the overall distance scale.
The pipeline  also uses a local reliability  metric that is a function of the neighbor number $k$, referred to as the \textbf{Trustworthiness Score}  $T(k)$.
 It examines how many of the nearest neighbors in the low-dimensional space are not neighbors in the original high-dimensional space. A higher   $T(k)$ (e.g., closer to 1) indicates  a better precision of local neighborhood preservation, thus fewer false neighbors accordingly.
Additionally, we also assess the computed embeddings by examining their cluster structure through the Silhouette Score and their outlier properties through LOF-related statistics.
In Appendix~\ref{app:metrics}, we provide a detailed description of the used scores along with their mathematical formulations.

\section{Hyperparameter Optimization}

There exists a broad family of DR algorithms that can be used to generate visualization plots for high-dimensional data.
In practice, the produced visualization plots are often sensitive to the hyperparameter configurations of the DR algorithm used.

For example, representative hyperparameters of the linear algorithm PCA include the number of principal components and the solver to use.
The  component number  is  set to 2 if PCA is used to directly generate a 2D visualization.
If it is used as a pre-processing step of another DR algorithm such as t-SNE or UMAP for denoising purposes, a higher component number is allowed, though associated with a computational cost trade-off. 
The solver is typically based on singular value decomposition (SVD),  usually  including the ``full'' option that leads to the exact SVD with a higher complexity, and the  ``randomized'' option that uses an approximated and computationally cheaper method.

For non-linear algorithms such as t-SNE and UMAP, visual artifacts and distortions of global spatial relationships are more sensitive to their hyperparameters. 
Example  hyperparameters of t-SNE include perplexity and learning rate.
Perplexity is a key parameter with a strong impact on the visualization output. 
It sets the target number of effective nearest neighbors for each data point,   balancing the algorithm's  attention between local and global  structures. 
 Lower values (e.g., $< 5$) usually result  in fragmenting continuous manifolds into isolated and noisy clusters, while higher values (e.g., $> 50$) focus  on global structure while merging separate clusters into a single mass. 
The learning rate of t-SNE determines the  step size of its weight update during KL divergence minimization. 
A high rate might cause convergence failure,  resulting in the data collapsing into being nearly uniformly distributed over a ball.
An overly low rate might cause the points to  be compressed into a dense cloud with a few outliers, which could be a result of the algorithm getting stuck in a suboptimal local minimum before forming natural clusters, or not yet converging into a meaningful representation of the underlying cluster structure. 
Example hyperparameters of UMAP include the neighbor number and \texttt{min\_dist}. 
Similar to perplexity, the UMAP neighbor number constrains the size of local neighborhood.
Small values restrict UMAP to a localized view and preserve  fine-grained details and local neighborhoods in the data.
On the contrary, higher values focus on global data structure and better preserve the global skeleton of the manifold.
The choice of \texttt{min\_dist} controls the minimum allowed Euclidean distance between points in the low-dimensional space. 
Lower values (e.g., 0.1) allow points to pack densely, while higher values (e.g., 0.5) prevent tight packing and aim to preserve the broad and continuous global shape of the data.

Selecting an appropriate hyperparameter setting for a DR algorithm (deciding algorithm configuration) is a delicate art. 
When done properly, it helps the viewer form an accurate understanding of the data and inspires meaningful discoveries.
Traditionally, hyperparameter selection is formulated as an optimization problem.
It searches for the optimal values in a pre-identified hyperparameter space by minimizing (or maximizing) a designed objective function, which, for example, can be a composite score function linearly combining those reliability metrics introduced earlier.   
Let $\theta$ denote the vector of hyperparameter values, $H$ the hyperparameter space, $\mathbf{Z}(A,\theta)$ the 2D embeddings computed by a DR algorithm $A$, and $O_i(\cdot )$ a penalty score that evaluates the computed embeddings. 
Assuming that we consider a set of $L$  score functions covering different evaluation aspects, one way to formulate a hyperparameter optimization problem is 
\begin{equation}
    \theta_A^* = \arg\min_{\theta \in H } \sum_{i=1}^L w_iO_i(\mathbf{Z}(A,\theta)),
\end{equation}
where $0\leq w_i \leq 1$ denotes the assigned weight to a score, controlling its contribution to the optimization.

Implementing such  a hyperparameter optimization process requires prior knowledge of the choices of the score functions $\{O_i\}_{i=1}^L$ and weight assignment $\{w_i\}_{i=1}^L$, as well as a pre-identified hyperparameter space $H$ to ease the search challenge while not missing promising settings.
This either requires strong expert knowledge of DR algorithm behavior and downstream expectation of the visualization result,  or induces another level of optimization process to find the best configuration of $H$, $\{O_i\}_{i=1}^L$  and $\{w_i\}_{i=1}^L$.
This is challenging for both algorithm design engineers and downstream visualization users.
Therefore, it motivates us to design an agentic pipeline to facilitate this process. 
For any given DR algorithm $A$, the pipeline exploits an LLM agent to reason about algorithm behavior and user needs,  employ either implicit or explicit scoring schemes to guide the hyperparameter search, and rapidly find satisfactory hyperparameter setting in a few iterations.

\section{Agentic AI Pipeline Design}

Our method formalizes the visualization generation  process as an iterative refinement loop guided by an LLM agent. 
Instead of relying on manual trial-and-error hyperparameter selection for a DR algorithm, we propose an automated pipeline that takes multi-modal input and employs an LLM agent to act as the expert evaluator. 
The agent continuously refines the DR algorithm configuration by taking into account (i) the generated visualization plot, (ii)  the 2D embeddings used for producing the plot, and (iii) a collection of metrics and extracted quantities summarizing the  geometric and distributional information of the data. 

\begin{figure*}
\centerline{\includegraphics[width=\linewidth]{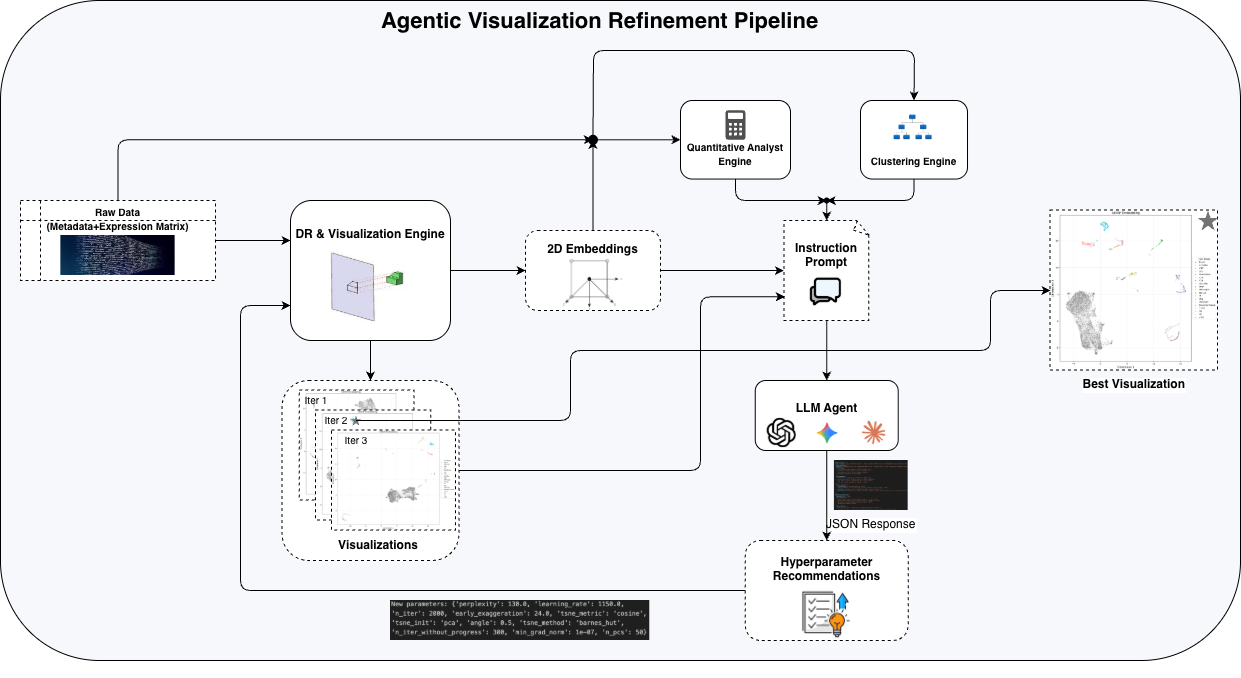}}
\caption{Overview of the Agentic Visualization Refinement Pipeline (AutoTuneVis). The system iteratively optimizes dimensionality reduction hyperparameters. Raw data is fed into the DR \& Visualization Engine to create 2D embeddings and scatter plots. The Quantitative Analyst and Clustering Engines extract metrics using these embeddings and raw data. These metrics and hierarchy dendrograms, alongside the plots, are combined into an Instruction Prompt for the LLM Agent. The agent evaluates the current state of the visualization and outputs a JSON response containing hyperparameter recommendations, which are fed back into the DR and Visualization Engine for the next iteration.  
The process continues until a stopping criterion is met, based on either an implicit quality score generated by the LLM agent or an explicit score whose weights are defined by the agent.}
\label{fig:agentic_pipeline}
\end{figure*}
\subsection{Iterative LLM-Guided Pipeline}
The pipeline consists of three phases, including visualization generation, multi-modal evaluation, and hyperparameter refinement, as illustrated in Figure \ref{fig:agentic_pipeline}.
In each iteration, a \textbf{DR \& visualization engine} computes 2D embeddings from the high-dimensional data  using the current DR configuration, and produces the corresponding visualization plot using the embeddings.
The embeddings are also processed by a \textbf{quantitative analyst engine} and a \textbf{clustering engine}.
The engine outputs, together with the embedding coordinates and visualization plot, are fed into the \textbf{LLM agent} to produce a structured JSON report that dictates the parameter search for the subsequent iteration. 
The loop terminates when the active score changes by less than 0.1 between consecutive iterations ($|s_t - s_{t-1}| < 0.1$), when the agent returns no actionable recommendation, or after a maximum of 10 iterations.
\subsection{Multi-Modal Input of LLM Agent }

To enable the LLM agent to act as a comprehensive visual diagnostician, we supply it with a rich context  comprising four distinct data modalities to reason over.

\noindent
 \textbf{Quantitative Metrics:} The quantitative analyst engine computes a set of scores as mentioned earlier, including   trustworthiness $T(k)$, Spearman $\rho$, Stress $S$, mean distance ratio $R$ and Silhouette scores, as well as LOF-based outlier statistics, to quantify embedding quality at both local and global scales.

\noindent
\textbf{Cluster Dendrograms:} The clustering engine produces two hierarchical clustering trees (dendrograms) using the original high-dimensional data  and the 2D embeddings, respectively. 
This allows the LLM agent to assess structural agreement and identify specific cluster (or sub-cluster) ordering.

\noindent
 \textbf{Embedding Coordinates:} The 2D embedding coordinates, along with summary quantities such as embedding statistics and centroid of labeled data groups when label information is available, are provided to the LLM agent. 
 This helps ground its spatial reasoning.

\noindent
\textbf{Visualization Plot:} The rendered visualization of the 2D embeddings  enables  the LLM agent to process Gestalt patterns, crowding, and visual artifacts that numerical metrics might fail to capture.
\subsection{Implicit and Explicit Scoring Schemes}  
Two distinct prompting strategies are developed for an LLM to evaluate visualization quality, guiding the generation of hyperparameter tuning trajectories.   
One is \textit{implicit scoring} that employs the LLM as a holistic visual diagnostician.  
It outputs a single, subjective quality score by independently judging the multi-modal input according to its own notion of overall quality. 
 The  other is \textit{explicit scoring} that employs the LLM   as a quantitative strategist. 
It designs a fixed weight vector $W = \{w_1, w_2, ..., w_n\}$ to formulate the composite objective function $O = \sum w_i O_i$ using quantitative metrics such as Trustworthiness, Spearman, and Stress. 
 The LLM recommends hyperparameters specifically to improve this composite objective function.
\subsection{Structured Output of JSON Diagnostics}
The LLM agent is prompted to produce a structured JSON report based on the multi-modal input. 
This JSON format ensures parsing reliability for the automated pipeline. 
The diagnostic JSON report includes the following information categories.

\noindent 
\textbf{Overall Assessment \& Quality Score:} The LLM agent is prompted to produce a high-level numerical score supported by rationale in text description.
The score is required to balance local preservation, global distortion, and cluster/class separation.

\noindent 
\textbf{Structural \& Visual Inspection:} The LLM agent is also prompted to provide a detailed comparison  between the dendrograms computed from the high-dimensional data and 2D embeddings, to identify   visual artifacts, e.g., crowding and artificial bridges, and to investigate domain-specific  patterns, e.g.,  biological cell-type relationships.
The findings are presented as  text description, which is also taken into account by the agent to form the quality score.

\noindent 
\textbf{Outlier Forensics:} An assessment of LOF distributions is requested, flagging potentially concerning patterns or data quality issues.

\noindent 
\textbf{Actionable Recommendation:} Most importantly, the LLM agent is prompted to produce a prioritized list of hyperparameter adjustments, e.g., the suggested setting of \texttt{n\_neighbors}, \texttt{min\_dist}, \texttt{n\_pcs} and other parameters for UMAP, alongside the rationale and expected impact in text description.
The suggested hyperparameter setting is to be used by the DR and visualization engine for computing 2D embeddings and producing the visualization plot in the next   iteration.

\section{LLM Input/Output Format}

\subsection{Input Format (State Representation)} 
At the beginning of each iteration, the pipeline compiles a ``master prompt'' containing the current algorithm parameters alongside an exhaustive suite of quantitative metrics. 
To convey hierarchical structures without visual plots, the agent is supplied with Newick tree string representations of both the high-dimensional data and the current 2D embedding. 
The input schema is structured as follows:

\begin{lstlisting}[basicstyle=\ttfamily\scriptsize, breaklines=true, breakatwhitespace=true, frame=single, caption={Schema of  Input Prompt of LLM}, label={lst:input_prompt}]
{
  "metrics": {
    "Global Structure & HD<->2D Correlation": {
      "Mean 2D/HD Distance Ratio": "0.6890",
      "Spearman Correlation": "0.7714",
      "Stress-1": "0.3673"
  },
    "Neighborhood Preservation": {
      "Trustworthiness (k=10)": "0.9547"
  },
    "Clustering & Label Agreement": {
      "Silhouette Score": "0.0414"
  },
    "Outlier Detection": {
      "LOF Median Score": "1.0123",
      "LOF Top 1% Score": "1.5468"
  },
    "Cluster Centroids (2D)": {
      "Centroid (Label Proximal tubule)": "(-6.1281, -0.9959)",
      "Centroid (Label NK cell)": "(20.9348, -1.7553)"
  },
    "Cluster Centroid Separation": {
      "Mean Centroid Distance": "17.6281",
      "Std Centroid Distance": "9.4125"
  },
  "label_summary": {
    "Proximal Tubule": 10392,
    "NK cell": 226, ...
  },
  "hierarchy_hd": {
    "newick": "((8,(10,(0,15))),...);",
    "dimension": "20D (PCA)"
  },
  "hierarchy_2d": {
    "newick": "((10,(0,15)),...);",
    "dimension": "2D"
  },
  "parameters": {
    "method": "tsne",
    "perplexity": 30.0,
    "learning_rate": 200.0, ...
  }
}
\end{lstlisting}

\subsection{Output Format (Diagnostic Response)} 

The LLM acts as an expert diagnostician, producing a detailed analysis report by reasoning about the input metrics and the visual plot (processed via vision capabilities). 
The agent is constrained to output a structured JSON response that articulates the  reasoning before making actionable hyperparameter recommendations. 
This structure allows the pipeline to parse the suggested parameters programmatically for the next iteration, and is detailed below.

\begin{lstlisting}[basicstyle=\ttfamily\scriptsize, breaklines=true, breakatwhitespace=true, frame=single, caption={Schema of   Diagnostic Output and Hyperparameter Recommendation.}, label={lst:output_response}]
{
  "quality_score": 6.0,
  "score_rationale": "Excellent local neighbor preservation but poor global structure...",
  "overall_assessment": {
    "key_strengths": ["High local fidelity"],
    "key_weaknesses": ["Global structure distortion"],
    "metric_analysis": { ... }
  },
  "dendrogram_comparison": {
    "agreement_level": "moderate",
    "key_differences": ["MNP relocated toward endothelial block..."]
  },
  "visual_inspection": {
    "artifacts": ["Large amorphous Proximal Tubule island"]
  },
  "recommendations": [
    {
      "parameter": "tsne.perplexity",
      "current_value": "30.0",
      "suggested_value": "80",
      "rationale": "Larger perplexity increases the effective neighborhood size...",
      "expected_impact": "Reduce Stress-1; more coherent macro-branches.",
      "priority": "high"
    },
    ...
  ],
  "follow_up_metrics": [ ... ]
}
\end{lstlisting}

Overall, by forcing the LLM to populate the \texttt{dendrogram\_comparison} and \texttt{visual\_inspection} fields before generating the \texttt{recommendations} array, the agent is guided to generate explanations tied directly to recommended actions. 
This ensures that the refinement process is transparent and enriches the context for further iterations.

\begin{figure*}[htbp]
    \centering
    \begin{minipage}[c]{0.45\textwidth}
        \begin{subfigure}[b]{\textwidth}
            \centering
            \includegraphics[width=\textwidth,height=0.16\textheight,keepaspectratio]{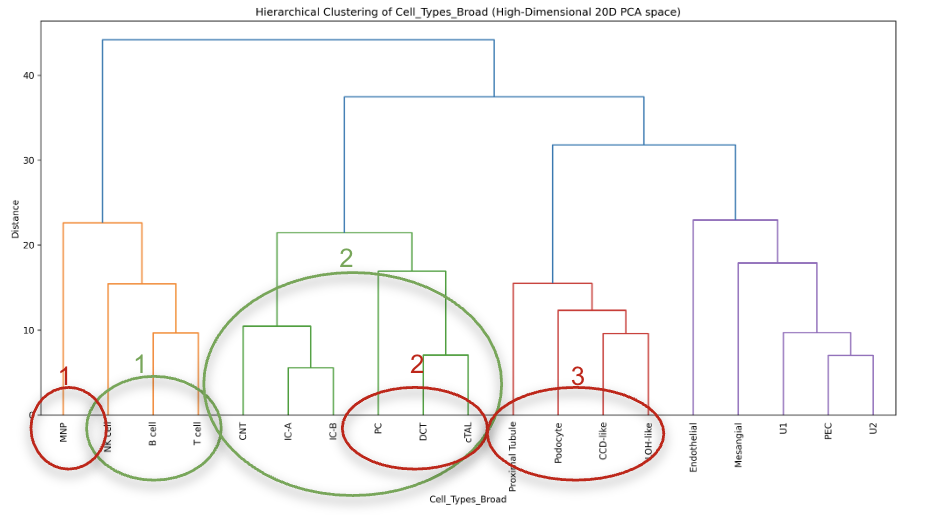}
            \caption{Dendrogram: High-Dimensional (200 PCA)}
            \label{fig:dendro_hd}
        \end{subfigure}
        
        \vspace{1.5em}
        
        \begin{subfigure}[b]{\textwidth}
            \centering
            \includegraphics[width=\textwidth,height=0.16\textheight,keepaspectratio]{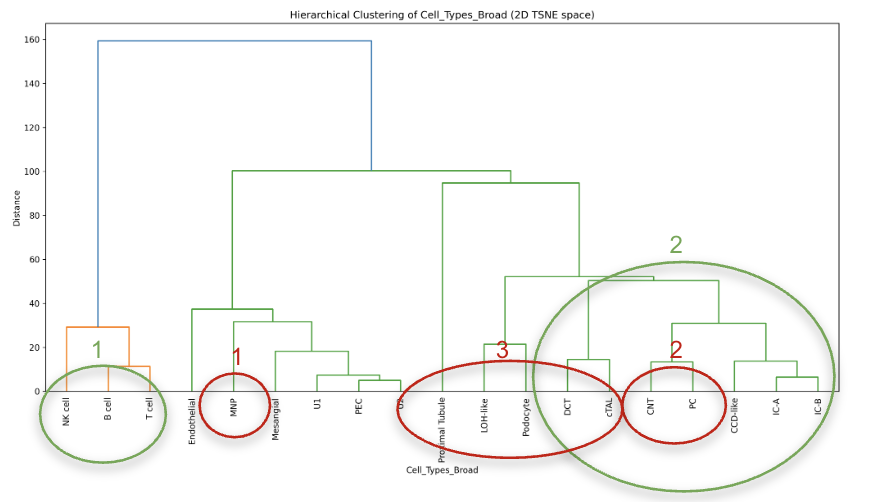}
            \caption{Dendrogram: Low-Dimensional (2D t-SNE)}
            \label{fig:dendro_ld}
        \end{subfigure}
    \end{minipage}
    \hfill
    \begin{minipage}[c]{0.52\textwidth}
        \begin{subfigure}[b]{\textwidth}
            \centering
            \includegraphics[width=\textwidth,height=0.34\textheight,keepaspectratio]{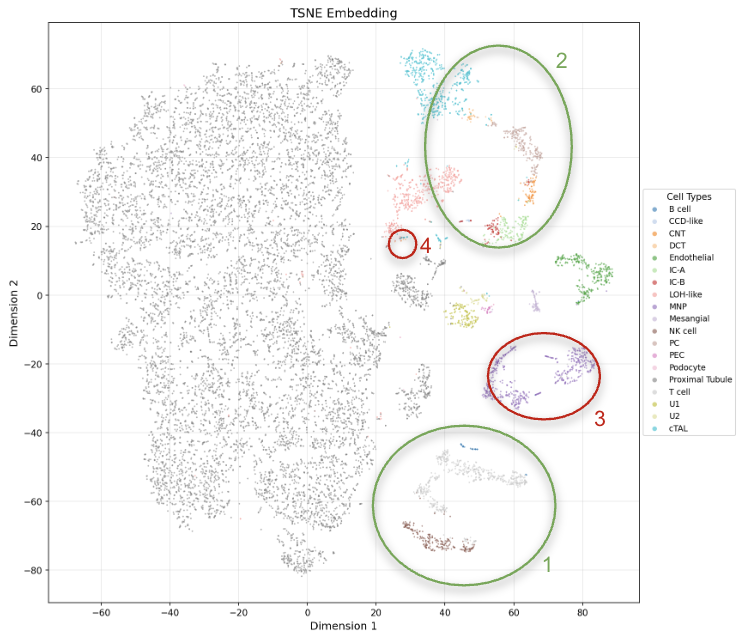}
            \caption{t-SNE Embedding Visualization}
            \label{fig:tsne_main}
        \end{subfigure}
    \end{minipage}

    \vspace{0.5em}
    \hrule
    \vspace{0.5em}

    \begin{minipage}[t]{0.48\textwidth}
        \footnotesize
        \textbf{Dendrogram Comparison}
        \begin{itemize}[leftmargin=*, noitemsep]
            \item \textbf{Agreement Level:} Moderate
        \end{itemize}

        \begin{itemize}[leftmargin=*, noitemsep]
            \item \textbf{Key Similarities:}
        \end{itemize}
        \begin{enumerate}[label=\arabic*., leftmargin=*, noitemsep]
            \item Immune grouping preserved: NK with (B, T) in both trees.
            \item Distal/collecting relationships appear in both.
        \end{enumerate}

        \begin{itemize}[leftmargin=*, noitemsep]
            \item \textbf{Key Differences:}
        \end{itemize}
        \begin{enumerate}[label=\arabic*., leftmargin=*, noitemsep]
            \item \textbf{MNP:} HD tree groups MNP with immune branch; 2D tree pulls MNP toward endothelial/mesangial block.
            \item \textbf{PC neighborhood:} HD links PC with DCT/CTAL; 2D places PC nearer CNT/ICs.
            \item \textbf{Proximal Tubule:} HD tree pairs PT with podocyte/LOH-like; 2D separates PT early.
        \end{enumerate}
    \end{minipage}
    \hfill
    \begin{minipage}[t]{0.48\textwidth}
        \footnotesize
        \textbf{Visual Inspection}
        \begin{itemize}[leftmargin=*, noitemsep]
            \item \textbf{Cluster Separation:} Good
            \item \textbf{Cluster Compactness:} Moderate
        \end{itemize}

        \begin{itemize}[leftmargin=*, noitemsep]
            \item \textbf{Notable Patterns:}
        \end{itemize}
        \begin{enumerate}[label=\arabic*., leftmargin=*, noitemsep]
            \item Immune triad (NK/B/T) cleanly separated.
            \item Distal nephron chain (DCT, cTAL, CNT, IC, PC) visible.
            \item MNP visually distant from other immune cells, likely artifact.
            \item Podocyte ($n=9$) forms a tiny island near LOH-like.
        \end{enumerate}
    \end{minipage}
    \vspace{0.5em}
    \hrule
    \vspace{0.5em}

    \caption{LLM agent (GPT-5.2) response and corresponding visual areas on cell type clustering: (a) High-Dimensional PCA dendrogram, (b) Low-Dimensional t-SNE dendrogram, and (c) the resulting t-SNE embedding on Healthy Human Kidney scRNA-seq  Dataset \citep{McEvoy2022} with Initial t-SNE Hyperparameter Setting. Enumerated green circles on dendrograms denote key similarities whilst enumerated red circles denote key differences. Circles on (c) denote notable patterns while green indicates positive and red indicates negative comments by the LLM agent.}
    \label{fig:case_initial}
\end{figure*}
\section{Experiments and Results}

To validate AutoTuneVis, we apply it to visualize  high-dimensional biological datasets on single-cell RNA sequencing (scRNA-seq) of kidney cells. 
The studied data patterns are particularly challenging due to severe class imbalance and the biological requirement to preserve both distinct cell types (clustering) and developmental trajectories (continuums).
The framework is tested on  multiple DR algorithms and state-of-the-art LLMs  based on two scoring schemes (implicit vs. explicit). 
We summarize the key experimental settings below and provide further details in Appendix~\ref{app:experiments}.

\noindent
\textbf{Datasets:} We visualize two datasets, including \textit{Healthy Human Kidney (Sex-Based Profiling)}   \citep{McEvoy2022} and \textit{Mature Human Kidney (Immune Zonation)}  \citep{Stewart2019}.

\noindent
\textbf{DR Algorithms:} The framework is tested on three non-linear DR algorithms including UMAP, t-SNE, and PaCMAP. 
Each algorithm presents a fundamentally different hyperparameter search space, allowing us to test the agent's adaptability.

\noindent
\textbf{LLM Agents:} To understand the impact of LLM versions, we compare three frontier LLMs, including GPT-5.2, Gemini-3.1-Pro-Preview, and Claude-Opus-4.5, for building the agentic pipeline.

\noindent
\textbf{Rendering and Schema Specification:}
All rendering parameters (e.g., resolution, axes, colormap, legend) are fixed across DR methods, datasets and providers.
The placeholders in Listing~\ref{lst:input_prompt} represent variable components, while the rest of the schema remains fixed.

\subsection{A Case Study for t-SNE Visualization}

\begin{table*}[htbp]
\centering
\caption{Performance across LLM agents, DR methods and scoring schemes. Mean $\pm$ std. over 3 independent runs per setting. Scores out of 10; higher is better.}
\label{tab:optimization_results}
\resizebox{\ifdim\width>\linewidth\linewidth\else\width\fi}{!}{
\begin{tabular}{@{}lllcccc@{}}
\toprule
\textbf{Dataset} & \textbf{DR Method} & \textbf{LLM} & \textbf{Implicit init.} & \textbf{Implicit best} & \textbf{Explicit init.} & \textbf{Explicit best} \\ \midrule
\multirow{9}{*}{\begin{tabular}[c]{@{}l@{}}Healthy \\  Human Kidney\\ 13.8k cells\end{tabular}}
 & \multirow{3}{*}{UMAP}   & GPT-5.2         & 8.03 $\pm$ 0.29 & 9.10 $\pm$ 0.10 & 8.15 $\pm$ 0.00 & 8.76 $\pm$ 0.10 \\
 &                         & Gemini-3.1-Pro    & 7.77 $\pm$ 0.46 & 9.00 $\pm$ 0.56 & 8.31 $\pm$ 0.00 & 8.67 $\pm$ 0.19 \\
 &                         & Claude-Opus-4.5 & 7.20 $\pm$ 0.00 & 7.20 $\pm$ 0.00 & 8.23 $\pm$ 0.00 & 8.25 $\pm$ 0.03 \\ \cmidrule(l){2-7}
 & \multirow{3}{*}{t-SNE}  & GPT-5.2         & 6.80 $\pm$ 0.40 & 9.00 $\pm$ 0.36 & 6.80 $\pm$ 0.00 & 8.36 $\pm$ 0.12 \\
 &                         & Gemini-3.1-Pro    & 5.99 $\pm$ 1.31 & 8.57 $\pm$ 0.06 & 7.02 $\pm$ 0.00 & 8.49 $\pm$ 0.10 \\
 &                         & Claude-Opus-4.5 & 5.60 $\pm$ 0.17 & 6.40 $\pm$ 0.85 & 7.06 $\pm$ 0.00 & 7.60 $\pm$ 0.54 \\ \cmidrule(l){2-7}
 & \multirow{3}{*}{PaCMAP} & GPT-5.2         & 8.30 $\pm$ 0.26 & 8.50 $\pm$ 0.46 & 8.40 $\pm$ 0.00 & 8.47 $\pm$ 0.07 \\
 &                         & Gemini-3.1-Pro    & 8.50 $\pm$ 0.00 & 8.63 $\pm$ 0.15 & 8.53 $\pm$ 0.00 & 8.65 $\pm$ 0.16 \\
 &                         & Claude-Opus-4.5 & 7.50 $\pm$ 0.00 & 7.50 $\pm$ 0.00 & 8.52 $\pm$ 0.00 & 8.52 $\pm$ 0.00 \\ \midrule
\multirow{9}{*}{\begin{tabular}[c]{@{}l@{}}Mature \\ Human Kidney\\ 20.1k cells\end{tabular}}
 & \multirow{3}{*}{UMAP}   & GPT-5.2         & 6.57 $\pm$ 0.25 & 7.33 $\pm$ 0.42 & 6.83 $\pm$ 0.00 & 7.16 $\pm$ 0.07 \\
 &                         & Gemini-3.1-Pro    & 6.77 $\pm$ 0.25 & 7.43 $\pm$ 0.40 & 7.08 $\pm$ 0.00 & 7.46 $\pm$ 0.17 \\
 &                         & Claude-Opus-4.5 & 5.80 $\pm$ 0.00 & 6.27 $\pm$ 0.50 & 7.08 $\pm$ 0.00 & 7.24 $\pm$ 0.14 \\ \cmidrule(l){2-7}
 & \multirow{3}{*}{t-SNE}  & GPT-5.2         & 6.50 $\pm$ 0.52 & 7.13 $\pm$ 0.45 & 7.36 $\pm$ 0.00 & 7.38 $\pm$ 0.02 \\
 &                         & Gemini-3.1-Pro    & 6.17 $\pm$ 0.58 & 7.27 $\pm$ 0.25 & 7.63 $\pm$ 0.00 & 7.80 $\pm$ 0.16 \\
 &                         & Claude-Opus-4.5 & 6.20 $\pm$ 0.00 & 6.20 $\pm$ 0.00 & 7.37 $\pm$ 0.00 & 7.37 $\pm$ 0.00 \\ \cmidrule(l){2-7}
 & \multirow{3}{*}{PaCMAP} & GPT-5.2         & 6.70 $\pm$ 0.10 & 6.97 $\pm$ 0.38 & 7.41 $\pm$ 0.00 & 7.41 $\pm$ 0.00 \\
 &                         & Gemini-3.1-Pro    & 7.23 $\pm$ 0.64 & 7.23 $\pm$ 0.64 & 7.66 $\pm$ 0.00 & 7.66 $\pm$ 0.00 \\
 &                         & Claude-Opus-4.5 & 6.60 $\pm$ 0.17 & 6.77 $\pm$ 0.25 & 7.50 $\pm$ 0.00 & 7.50 $\pm$ 0.00 \\ \bottomrule
\end{tabular}
}
%
%
\end{table*}

We exemplify the use of AutoTuneVis by testing it on t-SNE   to visualize the healthy human kidney data \citep{McEvoy2022}.
An example of the original response file output by the LLM is provided in Appendix~\ref{app:output}.

The pipeline is initiated with  a baseline t-SNE configuration, e.g., $\texttt{perplexity}=30$, $\texttt{learning\_rate}=200$, $\texttt{n\_pcs}=20$. 
The LLM assigns this initial embedding an implicit   quality score of 6.8, and generates a comprehensive JSON diagnostic report.
Figure~\ref{fig:case_initial} presents the initial visualization result together with a summary of the diagnostic report.
In this case, the baseline configuration successfully preserves local neighborhoods, evidenced by an exceptional Trustworthiness score of 0.9952. 
However, the LLM identifies fairly high global distortion and weak label separation, reflected in a Stress score of 0.5863, a moderate Spearman correlation of 0.5564, and a low Silhouette score of 0.0556.
After one iteration, the LLM settles on $\texttt{perplexity}=80$, $\texttt{learning\_rate}=800$, $\texttt{n\_iter}=3000$ and $\texttt{n\_pcs}=50$, raising the implicit score from 6.8 to 8.9. 
Below we summarize the LLM response on its diagnosis and hyperparameter refinement.

\vspace{1em}
\noindent
\textbf{LLM Diagnosis and Analysis}\\
Through the synthesis of dendrograms and visual metrics, the LLM identifies specific failure modes caused by data imbalance and algorithm settings. 
\textit{(i) Class Dominance:} The dataset suffers from extreme class imbalance, with the Proximal Tubule (PT) representing 75\% of cells. It forms a large, internally heterogeneous ``continent'' that compresses the dynamic range of the visual space and lowers label-based separation.
\textit{(ii) Sub-optimal Separability:} The Silhouette score is exceptionally low (0.0556), indicating substantial within-label dispersion and overlap between related tubular segments.
\textit{(iii) Structural Discrepancies:} While some local sub-trees are consistent, the LLM detects higher-level hierarchy distortions. For instance, the high-dimensional   dendrogram groups MNP cells coherently with the immune branch, whereas the 2D layout places them in a different branch. Furthermore, LOH-like and Endothelial populations show distorted topologies compared to the original  space.
\textit{(iv) Visual Artifacts:} The inspection notes thin and stringy  structures among immune and collecting-related groups, suggesting under-converged optimization or aggressive Barnes-Hut approximations.

\vspace{1em}
\noindent
\textbf{Hyperparameter Refinement Strategy}\\
Based on these diagnostics, the LLM outputs a prioritized list of specific hyperparameter adjustments to guide the next pipeline iteration. Here is an example.
\textit{(i) Increase Perplexity} ($\texttt{perplexity}: 30 \to 80$). Assigned a high priority, this change is recommended to increase the effective neighborhood size, improving global organization and reducing the fragmentation of gradients and higher-level relationships.
\textit{(ii) Increase Iterations} ($\texttt{n\_iter}: 1000 \to 3000$). Recommended to address the elongated ``threads'' and ensure complete convergence for a dataset of this size and complexity, leading to more stable islands.
\textit{(iii) Increase Learning Rate} ($\texttt{learning\_rate}: 200 \to 800$). Suggested to accelerate clean separation, reduce optimization artifacts, and prevent the algorithm from settling into crowded, stringy layouts.
\textit{(iv) Increase Principal Components} ($\texttt{n\_pcs}: 20 \to 50$). Assigned to better capture the subtle but true transcriptomic variation before applying t-SNE,   aiming to resolve the internal states of the dominant PT cluster.

\begin{figure*}[htbp]
    \centering
    \begin{minipage}[c]{0.45\textwidth}
        \begin{subfigure}[b]{\textwidth}
            \centering
            \includegraphics[width=\textwidth,height=0.16\textheight,keepaspectratio]{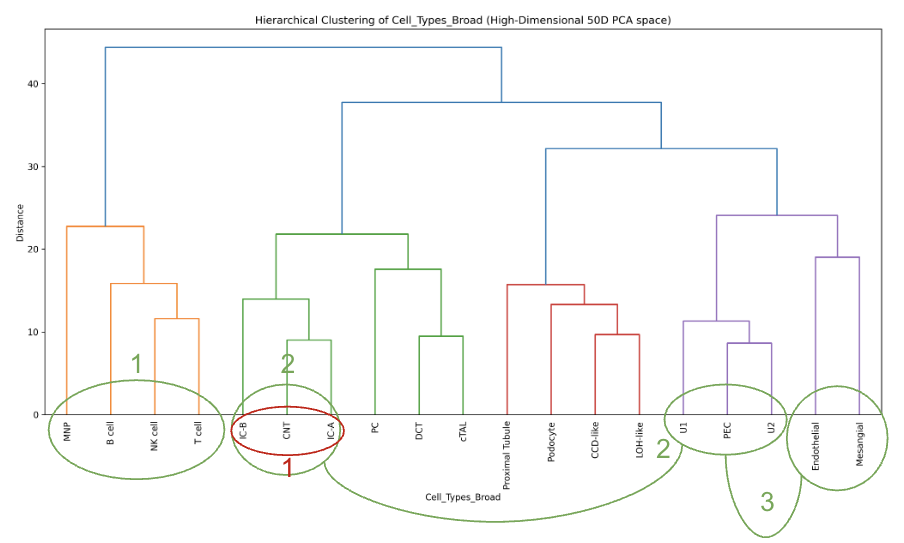}
            \caption{Dendrogram: High-Dimensional (200 PCA)}
            \label{fig:dendro_hd_opt}
        \end{subfigure}
        
        \vspace{1.5em}
        
        \begin{subfigure}[b]{\textwidth}
            \centering
            \includegraphics[width=\textwidth,height=0.16\textheight,keepaspectratio]{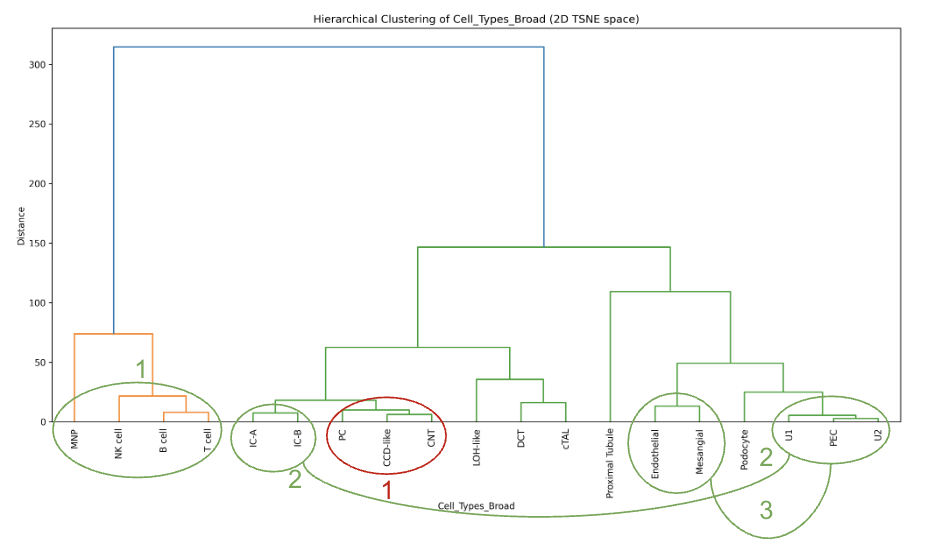}
            \caption{Dendrogram: Low-Dimensional (2D t-SNE)}
            \label{fig:dendro_ld_opt}
        \end{subfigure}
    \end{minipage}
    \hfill
    \begin{minipage}[c]{0.52\textwidth}
        \begin{subfigure}[b]{\textwidth}
            \centering
            \includegraphics[width=\textwidth,height=0.34\textheight,keepaspectratio]{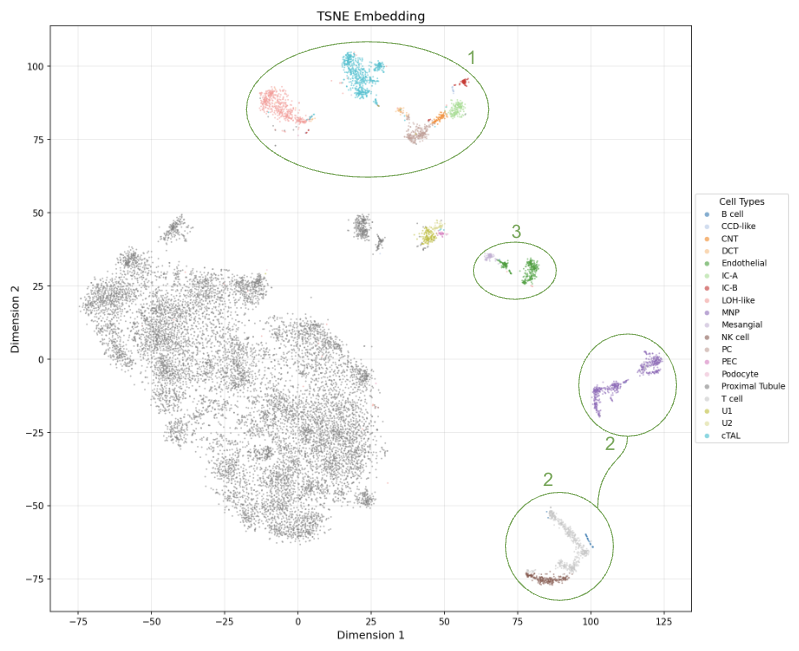}
            \caption{t-SNE Embedding Visualization}
            \label{fig:tsne_opt}
        \end{subfigure}
    \end{minipage}

    \vspace{0.5em}
    \hrule
    \vspace{0.5em}

    \begin{minipage}[t]{0.48\textwidth}
        \footnotesize
        \textbf{Dendrogram Comparison}
        \begin{itemize}[leftmargin=*, noitemsep]
            \item \textbf{Agreement Level:} High
        \end{itemize}

        \begin{itemize}[leftmargin=*, noitemsep]
            \item \textbf{Key Similarities:}
        \end{itemize}
        \begin{enumerate}[label=\arabic*., leftmargin=*, noitemsep]
            \item Top-level immune grouping is consistent: MNP clusters with (NK, (B,T)) in both 2D and PCA dendrograms (8 with (10,(0,15)) motif preserved).
            \item IC-A and IC-B remain a tight pair (5,6) and sit with PC/U1/U2/PEC-related branch similarly in both dendrograms.
            \item Endothelial with Mesangial (4,9) and U1/U2/PEC sub-branch (16,(12,17)) align between 2D and PCA, indicating good hierarchy preservation.
        \end{enumerate}

        \begin{itemize}[leftmargin=*, noitemsep]
            \item \textbf{Key Differences:}
        \end{itemize}
        \begin{enumerate}[label=\arabic*., leftmargin=*, noitemsep]
            \item Small rearrangements inside the distal/collecting subtree: 2D groups (11,(1,2)) earlier than PCA's ((6,(2,5))\ldots)---minor and likely within expected uncertainty for very small clusters.
            \item Leaf ordering differs locally but major clades are consistent.
        \end{enumerate}
    \end{minipage}
    \hfill
    \begin{minipage}[t]{0.48\textwidth}
        \footnotesize
        \textbf{Visual Inspection}
        \begin{itemize}[leftmargin=*, noitemsep]
            \item \textbf{Cluster Separation:} Excellent
            \item \textbf{Cluster Compactness:} Good
        \end{itemize}

        \begin{itemize}[leftmargin=*, noitemsep]
            \item \textbf{Notable Patterns:}
        \end{itemize}
        \begin{enumerate}[label=\arabic*., leftmargin=*, noitemsep]
            \item Clear distal nephron banding at the top: LOH-like, cTAL, DCT/CNT/CCD-like/IC-A/IC-B/PC cluster in a coherent neighborhood.
            \item Immune compartment is cleanly separated (B/T/NK with nearby MNP), aiding annotation QC.
            \item Vascular/stromal region (Endothelial, Mesangial, Podocyte, PEC) sits near each other as expected, with plausible proximities.
        \end{enumerate}
    \end{minipage}
    \vspace{0.5em}
    \hrule
    \vspace{0.5em}

    \caption{LLM   (GPT-5.2) response and corresponding visual areas on cell type clustering: (a) High-Dimensional PCA dendrogram, (b) Low-Dimensional t-SNE dendrogram, and (c) the resulting t-SNE embedding on Healthy Human Kidney scRNA-seq Dataset \citep{McEvoy2022} with Agent Optimized t-SNE Hyperparameter Setting. Enumerated green circles on dendrograms denote key similarities whilst enumerated red circles denote key differences. Circles on (c) denote notable patterns while green indicates positive and red indicates negative comments by the LLM agent.}
    \label{fig:case_optimized}
\end{figure*}

 To ensure that these parameter updates achieve the desired biological fidelity without degrading topology, the LLM dynamically generates follow-up metrics for the next iteration.
 These include tracking Trustworthiness at a larger neighborhood scale ($k=30$) to verify mid-range structure preservation, computing the Spearman correlation strictly within the PT cells to monitor internal gradient representation, and calculating the Silhouette score on finer subclusters to separate true biological mixing from label coarseness.
The final   visualization along with the LLM comments is presented in Figure~\ref{fig:case_optimized} with improved quality metrics.

\subsection{Evaluation of Hyperparameter Refinement}
\label{sec:eval}
\textbf{Multi-run Evaluation:} To reliably evaluate the hyperparameter tuning performance of an LLM agent, we independently repeat each tuning process 3 times for each dataset, driven by implicit scoring.
The initial and  best implicit scores achieved in each tuning process are used to compute the mean performance and standard deviation in the first two performance  columns of Table \ref{tab:optimization_results}. 
We also evaluate the initial and  best-tuned hyperparameters using an independent explicit composite score (a  weighted sum of DR metrics defined in Listing~\ref{lst:model_weights}), with the resulting scores reported in the last two  columns of Table \ref{tab:optimization_results}. 
The same initial setting is shared by all runs, therefore its corresponding explicit score has a standard deviation of zero.
Overall, Table \ref{tab:optimization_results} shows consistent performance improvements after hyperparameter tuning, where the low standard deviations indicate stable LLM behavior across runs.

\noindent
 \textbf{Human Validation:} To evaluate the LLM response,   4 human  experts on DR  independently assess each  sentence  in the  LLM-generated response files, assigning a score of 2 for ``fully correct and useful'', 1 for ``partially correct or of low relevance'', and 0 for ``incorrect or hallucinated''.
A total of 9 files, which correspond to optimization trajectories of 3 DR methods obtained by 3 LLMs, are assessed,  containing 2253, 290, and 290 statements for visualization diagnosis and analysis, hyperparameter recommendation, and refinement rationale, respectively.
Table~\ref{tab:external_runs} reports the per-trajectory results.  
Across the 9 optimization trajectories evaluated, 88.5\% of the 2,833 statements are fully correct. 
Accuracy is higher for hyperparameter recommendations and their rationales (91.7\% fully correct) than for visualization diagnosis and analysis (87.8\%).
The majority of the incorrect statements concern the comparison of the 2D and HD dendrograms.
Success in hyperparameter recommendations and  rationales indicates that the LLM is able to correctly identify the direction of hyperparameter adjustment. 
However, the responses show that the LLM operates at a relatively coarse granularity. For example, it can increase perplexity from 30 to 50 rather than making fine-grained adjustments, such as from 30 to 32.
Additionally, we  conduct another validation experiment, asking  human experts to independently assess visualization quality through inspecting   2D embedding plots. 
Further details are provided in Appendix~\ref{app:human}.

\begin{table}[htbp]
\centering
\caption{Human evaluation of LLM response per tuning trajectory on healthy human kidney dataset.}
\label{tab:external_runs}
\small
\resizebox{\ifdim\width>\linewidth\linewidth\else\width\fi}{!}{
\begin{tabular}{@{}lllrrrrrr@{}}
\toprule
\textbf{DR} & \textbf{LLM} & \textbf{Iter.} & \textbf{\# Sta. } & \textbf{Mean} & \textbf{2} & \textbf{1} & \textbf{0} \\ \midrule
t-SNE  & Gemini-3.1-Pro  & 4  & 207 & 1.82 & 88.4\% & 5.3\%  & 6.3\% \\
t-SNE  & Claude-Opus-4.5     & 5  & 292 & 1.79 & 81.8\% & 15.1\% & 3.1\% \\
t-SNE  & GPT-5.2             & 6  & 274 & 1.94 & 94.5\% & 5.1\%  & 0.4\% \\
UMAP   & Claude-Opus-4.5     & 5  & 271 & 1.82 & 86.0\% & 10.3\% & 3.7\% \\
UMAP   & Gemini-3.1-Pro      & 5  & 204 & 1.75 & 80.9\% & 12.7\% & 6.4\% \\
UMAP   & GPT-5.2             & 10 & 495 & 1.92 & 92.5\% & 7.1\%  & 0.4\% \\
PaCMAP & Claude-Opus-4.5     & 4  & 196 & 1.83 & 85.7\% & 11.7\% & 2.6\% \\
PaCMAP & Gemini-3.1-Pro      & 10 & 400 & 1.78 & 83.2\% & 12.0\% & 4.8\% \\
PaCMAP & GPT-5.2             & 10 & 494 & 1.95 & 94.7\% & 5.1\%  & 0.2\% \\ \midrule
\textbf{All} & & \textbf{59} & \textbf{2833} & \textbf{1.86} & \textbf{88.5\%} & \textbf{9.0\%} & \textbf{2.5\%} \\
\bottomrule
\end{tabular}
}
\end{table}
\noindent
 \textbf{Baseline Comparison:} We compare against the standard hyperparameter tuning approach of random search under a matched tuning budget. 
Grid search over 8-11 hyperparameters and meaningful Bayesian optimization are  infeasible within such a budget.
For LLM-tuning with implicit scoring,  each tuning process is repeated 3 times, with mean performance and standard deviation reported in  Table \ref{tab:random_search}, where LLM and random search perform similarly, each winning 3 of the 6 comparisons,  with a maximum performance difference of 0.46. 
In most cases, LLM-tuning terminates after fewer embedding evaluations (5.2 versus 10 on average) and therefore uses less wall-clock time on average (15.2 versus 16.4 minutes).
Overall, LLM  is able to   identify effective hyperparameter configurations more quickly than random search.
Moreover, it does not   require a predefined parameter range and  provides  a rationale for each tuning action.  

\begin{table}[htbp]
\centering
\caption{ Comparison with the random search baseline: explicit score at the best iteration, number of iterations run by the LLM loop, and total wall-clock time per run; mean $\pm$ sd over 3 runs. Initial is the default-parameter embedding (iteration 1, identical for both). }
\label{tab:random_search}
\small
\resizebox{\ifdim\width>\linewidth\linewidth\else\width\fi}{!}{
\begin{tabular}{@{}llcccccc@{}}
\toprule
 &  & \multicolumn{3}{c}{\textbf{Explicit score}} & \textbf{Iterations} & \multicolumn{2}{c}{\textbf{Time (min)}} \\
\cmidrule(lr){3-5} \cmidrule(lr){6-6} \cmidrule(lr){7-8}
\textbf{Dataset} & \textbf{DR} & \textbf{Initial} & \textbf{Random} & \textbf{LLM} & \textbf{LLM} & \textbf{Random} & \textbf{LLM} \\ \midrule
\multirow{3}{*}{Healthy} & UMAP   & 8.314 & 8.374 $\pm$ 0.102 & \textbf{8.667} $\pm$ 0.191 & 7.0 $\pm$ 4.4 & 12.0 $\pm$ 2.0 & 13.7 $\pm$ 7.5 \\
                       & t-SNE  & 7.021 & 7.962 $\pm$ 0.085 & \textbf{8.487} $\pm$ 0.102 & 4.3 $\pm$ 1.2 & 17.2 $\pm$ 5.4 & 22.1 $\pm$ 13.4 \\
                       & PaCMAP & 8.504 & \textbf{8.836} $\pm$ 0.085 & 8.651 $\pm$ 0.164 & 2.3 $\pm$ 0.6 & 6.3 $\pm$ 0.8 & 2.9 $\pm$ 0.9 \\ \midrule
\multirow{3}{*}{Mature} & UMAP   & 7.076 & \textbf{7.633} $\pm$ 0.071 & 7.461 $\pm$ 0.169 & 4.7 $\pm$ 2.1 & 18.5 $\pm$ 4.7 & 13.2 $\pm$ 8.2 \\
                       & t-SNE  & 7.631 & \textbf{7.816} $\pm$ 0.078 & 7.799 $\pm$ 0.156 & 4.3 $\pm$ 1.5 & 32.1 $\pm$ 14.2 & 23.2 $\pm$ 15.7 \\
                       & PaCMAP & 7.540 & 7.607 $\pm$ 0.110 & \textbf{7.662} $\pm$ 0.000 & 8.7 $\pm$ 2.3 & 12.2 $\pm$ 1.6 & 15.9 $\pm$ 3.9 \\ \bottomrule
\end{tabular}
}
\end{table}

\subsection{Ablation Studies}
\label{sec:ablation}
We ablate each multi-modal input component individually (by removing each component one-by-one), using Gemini-3.1-Pro-Preview to tune UMAP hyperparameters  on the healthy kidney dataset. 
Each tuning process is repeated 3 times.   Table \ref{tab:input_ablation} reports the performance.
We assess the effect of each component on both the diagnosis of the initial hyperparameters measured by implicit score estimation, and the quality of the best and final hyperparameters obtained after  tuning.
Removing each component lowers the best implicit score, except for the 2D dendrogram.
Removing the iteration history produces the largest drop and is the only ablation that also lowers the final score.
Human validation  identifies that incorrect statements are concentrated on the comparison of 2D and HD dendrograms, which limits the contribution of the hierarchy dendrograms (see Appendix~\ref{app:human_analysis}).
Moreover, all input components affect the initial  implicit score  estimated by the LLM (``Impl. init.''), verifying their contributions to LLM diagnosis and analysis.

When preparing the input, the Silhouette score, per-label centroids, cell-type distribution and dendrograms require label information. 
Table~\ref{tab:input_ablation} shows that removing the cell-type distribution or either dendrogram has a limited effect on the tuned performance.
This indicates  that the framework can operate without label-dependent components.
Removing the metrics, in contrast, leaves the quantitative gains intact but makes the LLM-estimated implicit score less stable. 
We also ablate the JSON order and visualization specification to examine the LLM robustness to input variations,
by tuning UMAP with Gemini-3.1-Pro-Preview on the healthy kidney dataset, with a shuffled JSON key order or a 90-degree rotation of the embedding applied to every LLM input of the refinement loop (nine runs per condition).
Results are reported in Table~\ref{tab:robust_loop}, where no detectable score differences are observed.
In general, the LLM stops its hyperparameter refinement after 2 to 10 iterations.
We observe a slightly earlier termination on average under both perturbations.
With this sample size, we cannot determine whether this results from a system-level effect or from the refinement pipeline itself, which we note as a limitation.
Further ablation experiment details, results and analysis are provided in Appendix~\ref{app:ablation}.

\begin{table}[htbp]
\centering
\caption{Input  ablations;  mean $\pm$ sd over 3 runs.}
\label{tab:input_ablation}
\small
\resizebox{\ifdim\width>\linewidth\linewidth\else\width\fi}{!}{
\begin{tabular}{@{}lcccc@{}}
\toprule
\textbf{Input Component} & \textbf{Impl. init.} & \textbf{Impl. best} & \textbf{Expl. best} & \textbf{Impl. final} \\ \midrule
Full Input         & 7.77 $\pm$ 0.46 & 9.00 $\pm$ 0.56 & 8.67 $\pm$ 0.19 & 7.97 $\pm$ 0.84 \\
\midrule
w/o quantitative metrics  & 6.50 $\pm$ 0.87 & 8.87 $\pm$ 1.18 & 8.68 $\pm$ 0.14 & 8.03 $\pm$ 2.63 \\
w/o embedding image       & 8.30 $\pm$ 0.00 & 8.43 $\pm$ 0.15 & 8.43 $\pm$ 0.13 & 8.39 $\pm$ 0.12 \\
w/o iteration history     & 7.93 $\pm$ 0.40 & 8.00 $\pm$ 0.44 & 8.31 $\pm$ 0.00 & 7.70 $\pm$ 0.35 \\
w/o 2D dendrogram        & 7.67 $\pm$ 0.29 & 9.23 $\pm$ 0.25 & 8.88 $\pm$ 0.02 & 8.93 $\pm$ 0.31 \\
w/o HD dendrogram  & 7.67 $\pm$ 0.29 & 8.87 $\pm$ 0.31 & 8.60 $\pm$ 0.08 & 8.67 $\pm$ 0.15 \\
w/o cell-type distribution& 7.93 $\pm$ 0.40 & 8.80 $\pm$ 0.30 & 8.78 $\pm$ 0.10 & 8.53 $\pm$ 0.55 \\
w/o current parameters    & 7.93 $\pm$ 0.40 & 8.73 $\pm$ 0.40 & 8.62 $\pm$ 0.11 & 8.47 $\pm$ 0.35 \\ 
\bottomrule
\end{tabular}
}
\end{table}
\subsection{Other   Analysis}

We conduct more comparisons to further understand the LLM behavior in guiding hyperparameter tuning for visualization algorithms, and analyze its cost.
Appendix~\ref{app:results} includes extra analysis.

\noindent
\textbf{Composite Score Weighting:}
When prompted to act as quantitative strategists, the three LLMs design  noticeably different composite objective functions to evaluate the 2D embeddings. 
For instance, GPT-5.2 prioritizes local neighborhood preservation and cluster separation, assigning the highest weights to Trustworthiness (0.30) and the Silhouette Score (0.30), while heavily discounting outlier detection (LOF Median: 0.05). 
Claude-Opus-4.5 demonstrates a strong bias toward strict clustering and identifying density anomalies, assigning 0.35 to the Silhouette Score and 0.20 to LOF, but placing less emphasis on global geometric preservation (Stress-1: 0.10). 
In contrast, Gemini-3.1-Pro-Preview adopts the most balanced optimization strategy, distributing weights relatively evenly across local topology (Trustworthiness: 0.25), clustering (Silhouette: 0.25), and global structure metrics (Spearman: 0.20, Stress-1: 0.15, LOF: 0.15).

\begin{figure}[t]
    \centering
    \begin{subfigure}[b]{0.49\textwidth}
        \centering
        \includegraphics[width=\linewidth]{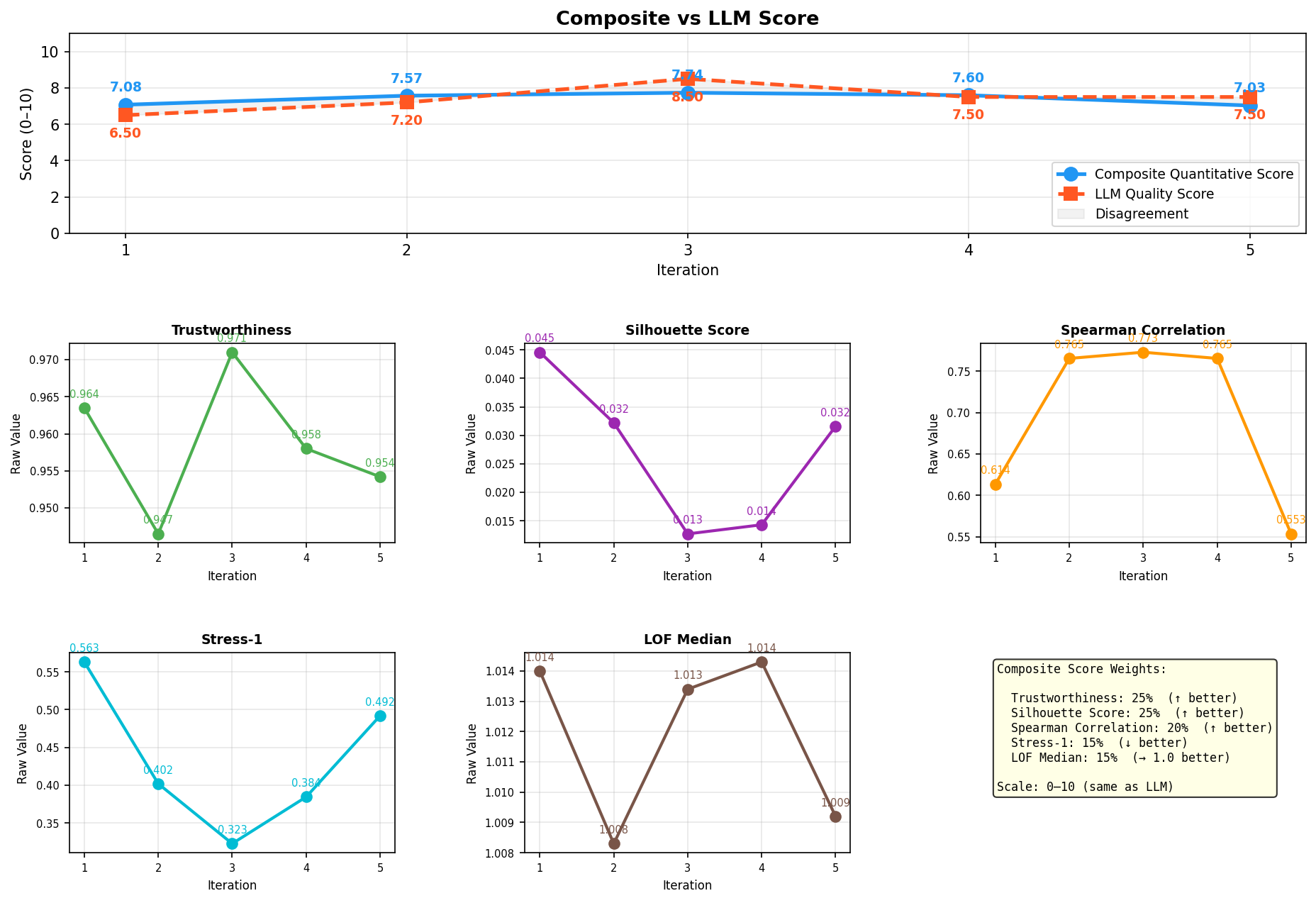}
        \caption{Gemini-3.1-Pro-Preview}
        \label{fig:traj_gemini}
    \end{subfigure}
    \hfill
    \begin{subfigure}[b]{0.49\textwidth}
        \centering
        \includegraphics[width=\linewidth]{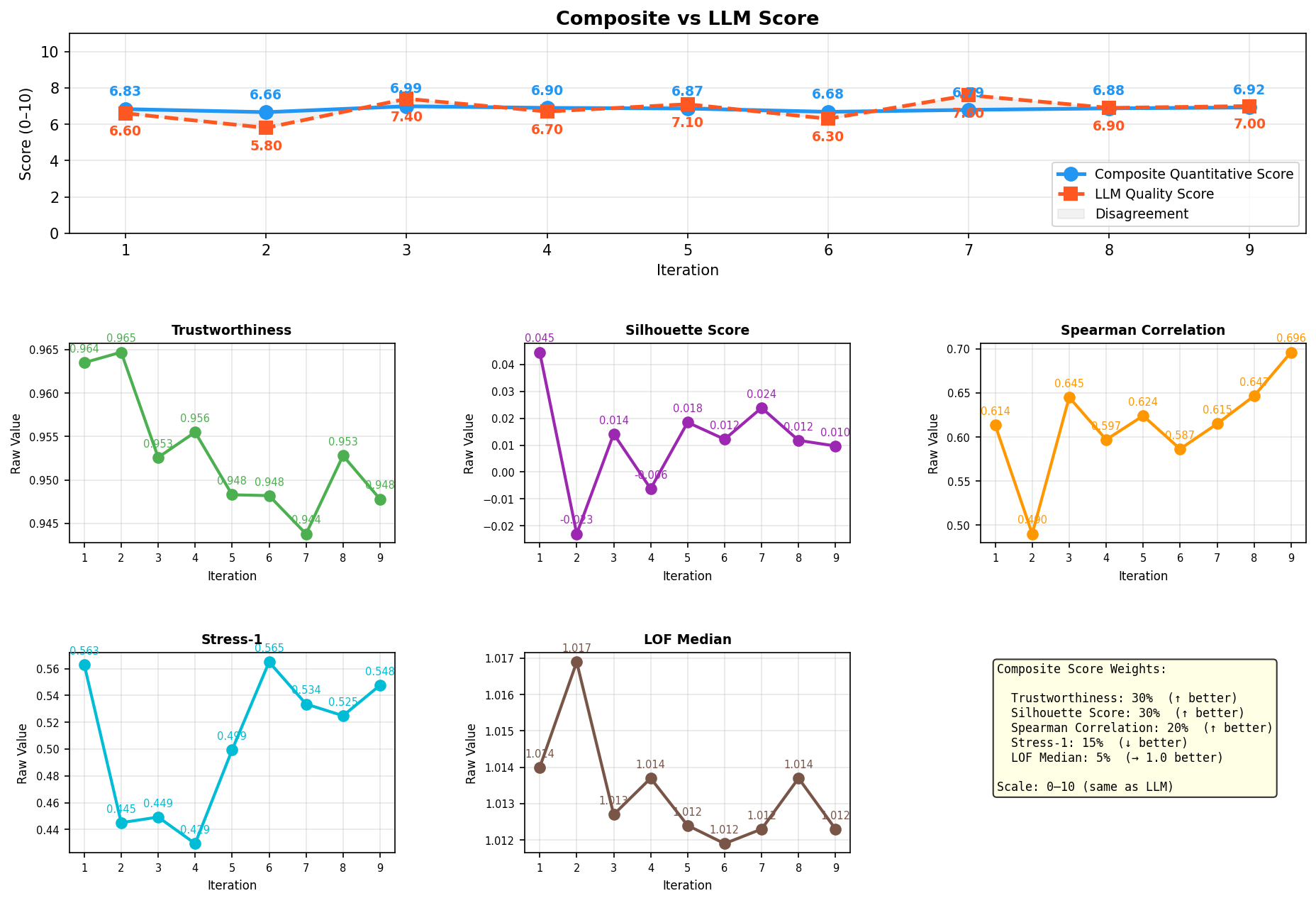}
        \caption{GPT-5.2}
        \label{fig:traj_gpt}
    \end{subfigure}

    \vspace{0.6em}

    \begin{subfigure}[b]{0.49\textwidth}
        \centering
        \includegraphics[width=\linewidth]{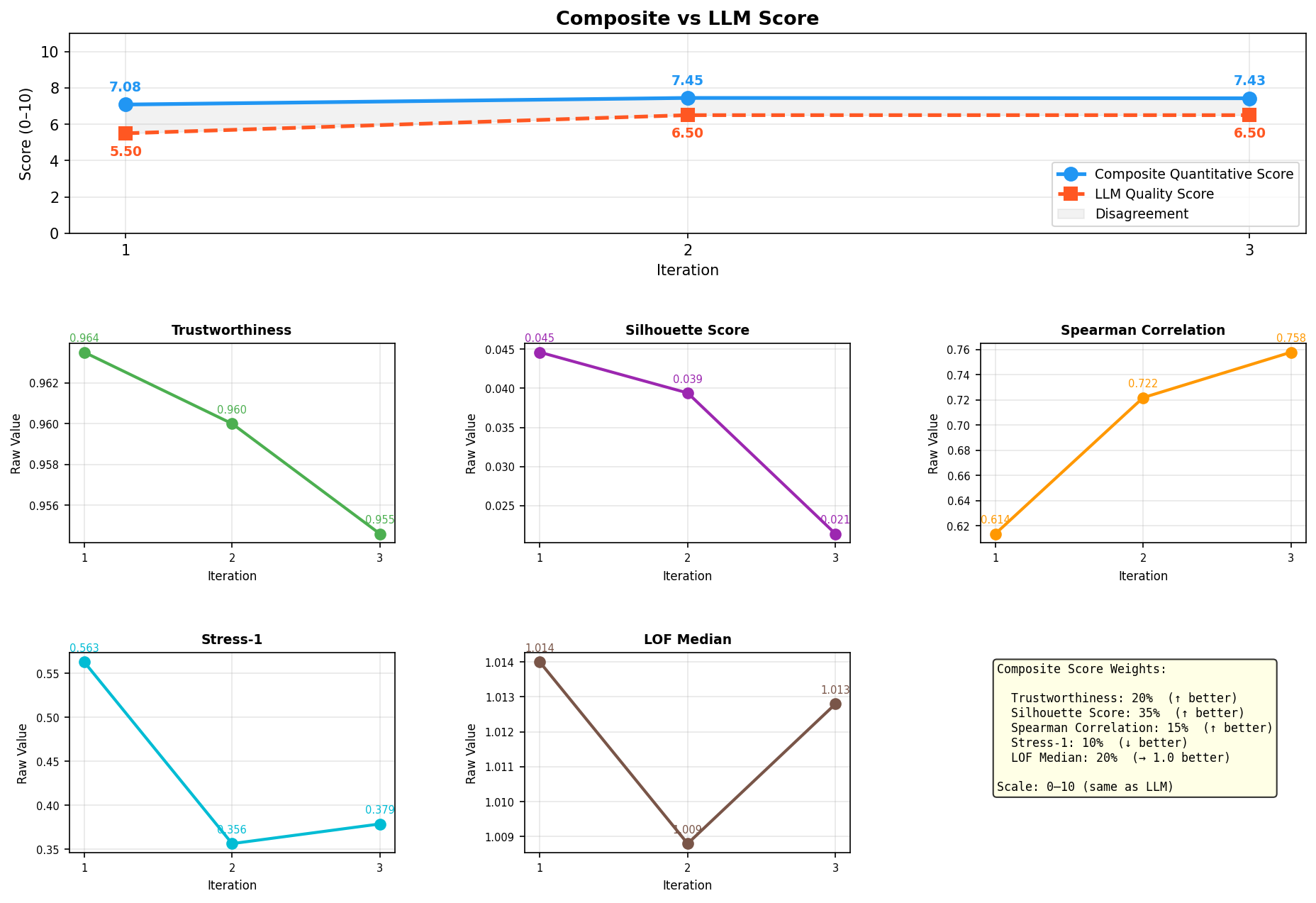}
        \caption{Claude-Opus-4.5}
        \label{fig:traj_claude}
    \end{subfigure}

    \caption{Comparison of hyperparameter tuning trajectories across LLM agents. Each plot illustrates the overlapping implicit and explicit score progressions alongside corresponding quantitative metrics over successive pipeline iterations.}
    \label{fig:optimization_trajectories}
\end{figure}

\noindent
\textbf{Tuning Trajectories and LLM Behaviors:} 
We illustrate the iterative hyperparameter tuning trajectories for Gemini-3.1-Pro-Preview, GPT-5.2, and Claude-Opus-4.5 in Figure \ref{fig:optimization_trajectories}.
By plotting both the implicit LLM-determined score and   explicit composite score alongside the raw metric values, optimization trade-offs and  distinct behavioral patterns emerge.
First of all, in almost all trajectories, improving global topology (reflected by decreasing Stress-1 and increasing Spearman correlation) comes at the direct expense of local structure (Trustworthiness and Silhouette Score). 
The three LLMs exhibit markedly different exploration strategies.
 Gemini-3.1-Pro-Preview stops in 5.2 iterations on average and reaches the highest best implicit score in the majority of settings.
GPT-5.2 runs the longest search, 8.4 iterations on average, with visibly more volatile metric trajectories, most notably in Stress-1 and Spearman correlation.
Claude-Opus-4.5 stops earliest, after 2.1 iterations on average. Its scores are also systematically lower than those of the other two providers for the identical embedding.
Overall, LLM agents typically identify their best hyperparameter configurations within the first few iterations (e.g., 2-4), with more iterations yielding little or no further performance improvement and sometimes beginning to cycle between previously tried values.
For instance, on the healthy human kidney dataset, 6 of the 9 implicit-scoring trajectories reach their best score within three iterations, and capping the loop at three iterations loses 0.17 points on average on the 0 to 10 scale while halving the number of iterations (59 to 27).

 In almost all cases, LLMs tune multiple hyperparameters per iteration rather than one, e.g., 3.1 on average for Gemini, 5.2 for Claude-Opus-4.5, and 5.5 for GPT-5.2.  
Although the LLM maintains search efficiency by  multi-parameter tuning,  it becomes difficult for the LLM to recognize the effect of each individual parameter.
We therefore record the complete parameter set per iteration in the history block, rather than only the changes, so that the agent can attribute observed effects to individual parameters.

\noindent
\textbf{Computational Cost Analysis:}  
The computational cost of hyperparameter tuning consists of   embedding computation, score evaluation, and one LLM call per iteration.
Embedding and score calculation can become computationally expensive for large-scale datasets,  primarily due to operations such as neighborhood graph construction  and the computation of Spearman correlation and Stress.
In our implementation, we mitigate  these costs using standard sub-sampling or approximate-kNN estimators.
With these approximations, across 54 experimental runs (Apple iMac, 8-core CPU, no GPU), the median per-iteration times for embedding and score evaluation are  115, 39,  and 264 seconds  on 13.8k cells and 158, 78, and 308 seconds on 20.1k cells,  for UMAP, PaCMAP, and t-SNE, respectively.
The median LLM-call latency per iteration is  38, 30, and 67 seconds for Gemini, GPT-5.2, and Claude-Opus-4.5, respectively,   requiring an average of  5.2, 8.4, and 2.1  refinement iterations.
Overall, a complete refinement takes 3-68 wall-clock minutes (median 17 minutes), with the long tail primarily arising from t-SNE runs that require  several thousand iterations.

\noindent
\textbf{API Cost Analysis:}  
The pipeline is suited to cost-sensitive deployments.
Because the LLM agent receives a fixed-size summary, the API cost per iteration is independent of data size; moreover, allowing more output tokens does not improve performance.  
In general, a complete refinement uses around 22k input  and 6k output tokens, corresponding to \$0.26 at   Opus-tier list price at the time of writing (US \$5 and \$25 per million input  and output tokens). 
Gemini and GPT runs have lower per-token costs.

\section{Conclusion}

Exploratory analysis and visualization of high-dimensional data is frequently bottlenecked by the subjective, labor-intensive process of hyperparameter tuning for DR algorithms. 
To tackle this, we have introduced a fully automated, agentic AI pipeline, AutoTuneVis, which leverages LLMs to refine and explain the visualization process.
By integrating quantitative reliability metrics, hierarchical dendrogram comparisons, and visual plots into a unified multi-modal context, the pipeline enables LLM agents to act as expert visual diagnosticians.
Bridging the gap between mathematical evaluation and qualitative  insight, it can rapidly navigate complex hyperparameter spaces across different DR algorithms (t-SNE, UMAP, and PaCMAP), and provide  a rationale for every parameter adjustment.

Experimenting on complex biological datasets,  we have quantitatively and qualitatively examined the validity of the LLM-generated responses and the quality of  visualization results based on both benchmark metrics and human validation, where $88.5\%$ of 2,833 LLM-generated statements are confirmed to be fully correct and relevant. 
This verifies the capacity of AutoTuneVis for supporting the production of scatter plots that faithfully represent the underlying high-dimensional reality, accelerating pattern discovery without the need for exhaustive manual tuning, and enhancing the understanding of DR algorithms and data behavior. 
Despite  efforts to ease the burden of hyperparameter tuning, the computational cost of classical embedding algorithms and their evaluation remains a bottleneck for  massive datasets, particularly due to neighborhood-graph-related operations. 
Our future work will explore the potential of  LLM-assisted algorithm design to accelerate these computationally intensive  operations.

\vspace{-6pt}
\acks{This work is supported by the Bicentenary PhD Studentship from the Faculty of Science and Engineering, University of Manchester.}

\newpage
\appendix

\section{Visualization Reliability Metrics}
\label{app:metrics}

\subsection{Global Reliability  Metrics}

\textbf{Spearman Correlation Score} \citep{spearman1904} computes the Spearman's rank correlation coefficient for distance ranks.
Specifically, we compute the rank difference by comparing  the    pairwise distance ranks  in the original high-dimensional  space and in the reduced low-dimensional   space, yielding
\begin{equation}
\rho_s = 1 - \frac{6 \sum_{i < j} \left(r_{ij}^{HD} - r_{ij}^{LD}\right)^2}{N_p(N_p^2 - 1)},
\end{equation}
where $r_{ij}^{HD}$ and $r_{ij}^{LD}$ denote the ranks of pairwise distances between points $i$ and $j$ in the high-dimensional and low-dimensional spaces, respectively, and $N_p$ is the number of ranked observations, i.e. the number of point pairs entering the correlation, rather than the number of data points. When all pairs of a dataset of $n$ points are used, $N_p = n(n-1)/2$; when the pairwise distances are estimated from a random subsample of pairs, $N_p$ is the number of sampled pairs.
A higher value of $\rho_s $, e.g., being closer to 1,  indicates that the relative ordering of the pairwise distances is overall well preserved for all the point pairs in the data.

\noindent
\textbf{Stress Score} \citep{venna2006local} quantifies  distance distortion from the high-dimensional space to the low-dimensional space. 
It measures the normalized residual between the high- and low-dimensional pairwise distances after min-max normalization of each distance set, i.e. the overall pairwise distance distortion, and it is given as
\begin{equation}
S =
\sqrt{
\frac{
\sum_{i<j} \left(d_{ij}^{LD} - d_{ij}^{HD}\right)^2
}{
\sum_{i<j} \left(d_{ij}^{HD}\right)^2
}},
\end{equation}
where $d_{ij}^{HD}$ and $d_{ij}^{LD}$ denote the normalized pairwise distances in the high-dimensional and low-dimensional spaces.
The stress score provides an overall examination of embedding fidelity. 
A lower value indicates a better preservation of the original data geometry.

\noindent
\textbf{Mean Distance Ratio} focuses on assessing the scaling change of pairwise distance from the high-dimensional space to low-dimensional space, computed by
\begin{equation}
R = \frac{1}{N_p} \sum_{i<j} \frac{d_{ij}^{LD}}{d_{ij}^{HD}},
\end{equation}
where $N_p = \frac{n(n-1)}{2}$ is the total number of unique point pairs. 
It examines the compression or expansion of the data points before and after DR, and is reported to the agent as an interpretable descriptor of global scaling.
A value of $R$ close to 1 indicates a better preservation of the overall distance scale.

\noindent
As a robustness check we also compute the median log ratio
$R_{\log} = \operatorname{median}_{i<j} \log (d_{ij}^{LD} / d_{ij}^{HD})$,
which is symmetric between expansion and compression and bounded against
outlying pairs. The two statistics agree on the direction of scale change
in 51 of the 54 runs from Table~\ref{tab:optimization_results}, so the conclusions drawn from $R$ are not artifacts
of its aggregation; the three disagreements are settings where $R$ lies
close to 1 and the direction is not meaningful in either statistic.

\subsection{Local Reliability  Metrics}

\textbf{Trustworthiness Score}  \citep{venna2006local} measures how many of the nearest neighbors in the low-dimensional space are  not neighbors in the original high-dimensional space, computed by
\begin{equation}
T(k) = 1 - \frac{2}{n k (2n - 3k - 1)} 
\sum_{i=1}^{n} \sum_{j \in U_i^k} \left(r_{i,j}^{HD} - k\right),
\end{equation}
where $U_i^k$ is the set of points that are among the $k$-nearest neighbors of $i$ in the low-dimensional space but not in the high-dimensional space, and $r_{i,j}^{HD}$ denotes the rank of the point $j$ in the high-dimensional ordering around $i$.
The score quantifies the amount of spurious local structure introduced by the embedding.
 A higher value, e.g., being closer to 1, indicates  a better precision of local neighborhood preservation, thus fewer false neighbors accordingly.

\subsection{Clustering and Label Agreement Metrics}
 
\textbf{Silhouette Score} \citep{rousseeuw1987} measures how well the low-dimensional embedding separates the given cell-type labels, computed by
\begin{equation}
\mathrm{Sil} = \frac{1}{n} \sum_{i=1}^{n} \frac{b_i - a_i}{\max(a_i, b_i)},
\end{equation}
where $a_i$ is the mean 2D distance from point $i$ to all other points with the same label, and $b_i$ is the smallest mean 2D distance from $i$ to the points of any other label.
The score lies between -1 and  1.
Values near 1 indicate compact, well-separated label clusters, values near 0 indicate overlapping clusters, and negative values indicate points closer to a foreign cluster than to their own.

\subsection{Outlier Metrics}
 
\textbf{Local Outlier Factor} (LOF) \citep{breunig2000lof} measures the density of each point relative to its $k$-nearest neighbors in the low-dimensional space, computed by
\begin{equation}
\mathrm{LOF}_k(i) = \frac{1}{k} \sum_{j \in N_k(i)} \frac{\mathrm{lrd}_k(j)}{\mathrm{lrd}_k(i)},
\end{equation}
with the local reachability density computed by
\begin{equation}
\mathrm{lrd}_k(i) = \left( \frac{1}{k} \sum_{j \in N_k(i)} \mathrm{rd}_k(i, j) \right)^{-1},
\end{equation}
where $N_k(i)$ is the $k$-nearest-neighbor set of $i$, $\mathrm{rd}_k(i,j) = \max\{d_k(j),\, d(i,j)\}$ is the reachability distance, and $d_k(j)$ is the distance from $j$ to its $k$-th nearest neighbor.
A value near 1 indicates that a point lies in a region of density comparable to its neighbors; values well above 1 indicate points that are sparser than their surroundings and are flagged as local outliers.

Two summaries of the LOF distribution are reported.
The \textbf{LOF median} $\mathrm{LOF}_{\mathrm{med}}$, i.e., the median over all points, stays near 1 when the embedding places most cells in density-consistent regions. 
The \textbf{LOF top 1\%} $\mathrm{LOF}_{1\%}$, i.e., the mean over the 1\% of points with the highest LOF,  characterizes the tail of the distribution.
A large value indicates isolated cells or fragmented clusters, whereas a value close to the median indicates a homogeneous density with no severe outliers.

\begin{table}
\caption{Summary of evaluation metrics and their interpretation.}
\small
\begin{tabular*}{\columnwidth}{@{}l@{\extracolsep{\fill}}lll@{}}
\toprule
Metric & Measures & Ideal & Aspect \\
\midrule
$\rho_s$ & Rank consistency & High ($\uparrow$) & Global order \\
$S$ & Metric distortion & Low ($\downarrow$) & Global geom. \\
$R$ & Scaling exp/comp & $\approx 1$ & Global scale \\
$T(k)$ & False neighbors & High ($\uparrow$) & Local prec. \\
$\mathrm{Sil}$ & Label separation & High ($\uparrow$) & Cluster quality \\
$\mathrm{LOF}_{\mathrm{med}}$ & Typical density & $\approx 1$ & Density consistency \\
$\mathrm{LOF}_{1\%}$ & Density tail & Low ($\downarrow$) & Outliers \\
\bottomrule
\end{tabular*}
\label{tab:metric_summary}
\end{table}

\section{More Details on Experiments}
\label{app:experiments}

\subsection{Dataset Information} 

We study two datasets. 
One is \textit{Healthy Human Kidney (Sex-Based Profiling)} sourced from the study ``Single-cell profiling of healthy human kidney reveals features of sex-based transcriptional programs and tissue-specific immunity'' \citep{McEvoy2022}.
This dataset  consists of cells with distinct type labels but suffers from extreme class imbalance; Proximal Tubule (PT) cells dominate approximately 75\% of the space. 
It presents challenges in preserving distinct, highly localized immune cell clusters alongside broader sex-based transcriptional differences.
The other dataset is \textit{Mature Human Kidney (Immune Zonation)} sourced from the study ``Spatio-temporal immune zonation of the human kidney'' \citep{Stewart2019}. 
It features complex continuous trajectories representing immune zonation, rigorously testing the pipeline's ability to preserve global manifold structures without artificial fragmentation. 
Both datasets are   downloaded from the UCSC Cell Browser \citep{speir2021ucsc}.

\subsection{Rendering and Schema Specification}

All rendering parameters are fixed across DR methods, datasets and providers.
The scatter plot is drawn at $12 \times 10$ inches and 150\,dpi with equal aspect ratio, with categorical colors from \texttt{tab20}, extended by \texttt{gist\_rainbow} when the number of labels exceeds twenty. 
The plot has a cell-type legend placed outside the plot area, no in-plot point labels, axis labels ``Dimension 1'' and ``Dimension 2'', and a light background grid. The image is submitted as a single PNG. 
The placeholders shown in Listing~\ref{lst:input_prompt} represent variable components, while the rest of the schema remains fixed.

\subsection {Neighbor Choice  and  Cluster  Granularity} 
All neighborhood-based measures reported to the agent use $k=10$, which is the value used throughout the experiments for trustworthiness and the $k$-NN based checks. 
The two dendrograms are not affected by the choice of $k$.
They are built by Ward linkage on the per-label centroids, so the number of leaves equals the number of annotated cell types and no cluster number needs to be fixed. 
The compact ``clusters at height $\leq 5$'' summary passed to the agent uses a fixed cut height in the respective space, applied identically to the high-dimensional and the two-dimensional tree so that the two are comparable.
Where a partition is required for a label-agreement measure, the number of clusters is set to the number of annotated labels.

\section{More Results and Analysis}
\label{app:results}

\subsection{Additional Human Validation Analysis}
\label{app:human_analysis}
We provide further analysis to complement the validation result of statements in the LLM-generated hyperparameter tuning trajectories reported in Section~\ref{sec:eval}. 
We notice that the majority of the incorrect statements concern the comparison of the 2D and HD dendrograms, either directly or through cell-type relationships inferred from them, e.g., reporting inconsistent placement of B cells or of IC-A/IC-B between 2D and HD when both are consistent. 
This agrees with the component-wise ablation study which shows that removing the 2D dendrogram field even has a positive effect, since information inferred from the dendrogram fields is often incorrect or prone to hallucination.
Metric interpretation, visual inspection and outlier analysis contain far fewer incorrect statements overall, and 3 out of 4 experts note that the agent draws attention to subtle details a human observer could overlook.
On parameter scale, all experts confirm  that the direction of every adjustment is correct but that the magnitudes are coarse, e.g., perplexity 30 to 100 rather than a fine step, and occasionally bold, e.g., learning rate 200 to 800.
The experts also conclude that the most useful recommendations mostly occur in the first two iterations, after which the agent reverts to previously tried values (perplexity $30 \rightarrow 100 \rightarrow 30 \rightarrow 130 \rightarrow 30$), and recommend  capping the loop at three iterations.

 \subsection{Additional Human Validation Experiment}
\label{app:human}
Another expert validation experiment is conducted to assess visualization quality independently of the agent's own score. 
For each of the 18 tuning settings,  4 plotted embeddings of the same data are shown to 4 human DR experts side by side without captions, produced using the default hyperparameters, the LLM-tuned hyperparameters, the random-searched  hyperparameters under a matched budget, and a corrupted embedding produced by plausible but mis-set hyperparameters, serving as a control.
The experts rank each embedding   from 1 (best) to 4 (worst) with ties allowed, examining (i) cluster separation, (ii) within-cluster coherence, (iii) multiscale structure, and (iv) global display. 
The score of an embedding is its mean rank over these four aspects.
Table~\ref{tab:blind_ranking} reports the results.
The corrupted embedding ranks last on every criterion in every setting, confirming that the ranking discriminates on embedding quality.
The LLM-tuned embedding obtains the best mean score (1.65), ahead of the default (1.79) and random search (2.33), and is preferred over random search in 14 of the 18 settings, tied in 3 and worse in 1.
The comparison with the default is method-dependent.
Specifically, t-SNE defaults rank  worst (2.38) and LLM tuning improves them to 1.46, winning 5 of 6 settings and losing none, whereas UMAP and PaCMAP defaults are already close to optimal (1.54, 1.46) and tuning leaves them essentially unchanged (1.88, 1.62).
The LLM-tuned embedding is   the most consistent across settings (sd 0.39 versus 0.62 for the default and 0.47 for random search), with a worst case of 2.25 against 3.00 for both alternatives.
Therefore, the benefit of tuning is better described as a floor on embedding quality than as a uniform improvement.

\begin{table}[htbp]
\centering
\caption{Blind expert ranking of embeddings over 18 tuning processes (1 = best, 4 = worst); mean rank reported.}
\label{tab:blind_ranking}
\small
\resizebox{\ifdim\width>\linewidth\linewidth\else\width\fi}{!}{
\begin{tabular}{@{}lcccc@{}}
\toprule
 & \textbf{Default} & \textbf{Random} & \textbf{LLM} & \textbf{Corrupted} \\ \midrule
\multicolumn{5}{@{}l}{\emph{All 18 settings}} \\
Cluster separation        & 1.67 & 2.67 & \textbf{1.44} & 4.00 \\
Within-cluster coherence  & 1.78 & 2.39 & \textbf{1.61} & 4.00 \\
Multiscale structure      & 1.94 & 2.06 & \textbf{1.78} & 4.00 \\
Global display            & 1.78 & 2.22 & \textbf{1.78} & 4.00 \\
Mean (all criteria)       & 1.79 & 2.33 & \textbf{1.65} & 4.00 \\ \midrule
\multicolumn{5}{@{}l}{\emph{By DR method (6 settings each)}} \\
t-SNE                     & 2.38 & 2.00 & \textbf{1.46} & 4.00 \\
UMAP                      & \textbf{1.54} & 2.42 & 1.88 & 4.00 \\
PaCMAP                    & \textbf{1.46} & 2.58 & 1.62 & 4.00 \\
\bottomrule
\end{tabular}
}
\end{table}

 \subsection{Additional Ablation Studies}
\label{app:ablation}
In these ablation experiments, for each input change, we first conduct a single-round experiment on UMAP with fixed hyperparameters for 3 independent runs (Table~\ref{tab:robust_single}), and then conduct a full-tuning-loop experiment that runs the whole hyperparameter refinement loop for UMAP  with Gemini-3.1-Pro-Preview on the healthy kidney dataset for 9 independent runs (Table~\ref{tab:robust_loop}).
Note that the \ref{tab:robust_single} and \ref{tab:robust_loop} and the ``Full Input'' setting in Table~\ref{tab:input_ablation} correspond to the same experimental configuration but to different sets of random runs.

\noindent
\textbf{JSON Ordering:} We examine the effect of JSON ordering in the LLM input, by shuffling the key order of every JSON block.
Results of the single-round experiment  (3 shuffles $\times$ 3 repeats $\times$ 3 providers) in Table~\ref{tab:robust_single} show no ordering effect beyond sampling noise.
The change in implicit score is at most 0.09 points and stays within the repeat-to-repeat spread of the unperturbed condition.
When the whole hyperparameter refinement loop is run with the shuffled order at every iteration, no detectable score difference is observed over 9 runs (Table~\ref{tab:robust_loop}).

\noindent
\textbf{Input Specification:} We ablate  the input by consistently applying a rigid transform (three rotations and two reflections, 5 variants) to the 2D embeddings and to the 2D centroids in the JSON so that plot and metadata remain in agreement. 
 Every judgment is a fresh, history-free request. 
In the single-round experiment, the implicit score changes by at most 0.19 points across all providers (Table~\ref{tab:robust_single}). 
In the full-tuning-loop experiment,  no detectable score difference is observed after the transformation either (Table~\ref{tab:robust_loop}).
We also observe that the hyperparameter refinement terminates earlier on average, i.e., from 8.0 $\pm$ 3.3 iterations to 6.2 $\pm$ 2.9 under JSON reordering and to 5.4 $\pm$ 3.0 under the transformation.
In practice, the refinement stops when two consecutive scores differ by less than 0.1, while the agent often oscillates between parameter settings and hence between scores.
Termination is therefore largely determined by the sampling variation of the LLM.  
Therefore, the iteration count is more a consequence of this system-level effect than an indicator of robustness.
Future work will explore a more effective and consistent use of iterations.

\begin{table}[htbp]
\centering
\caption{Robustness of the implicit score to input perturbations, single-round scoring of one fixed embedding (UMAP, healthy human kidney, initial configuration). Mean $\pm$ sd of the implicit score over repeats ($n=3$/$9$/$15$ per provider, by row); $\Delta$ is the change from the unperturbed condition of the same provider. JSON order: 3 shuffles; orientation: 3 rotations and 2 reflections, applied jointly to the plot and to the 2D centroids in the JSON. Every judgment is a fresh, history-free request.}
\label{tab:robust_single}
\small
\resizebox{\ifdim\width>\linewidth\linewidth\else\width\fi}{!}{
\begin{tabular}{@{}lcccccc@{}}
\toprule
 & \multicolumn{2}{c}{\textbf{Claude-Opus-4.5}} & \multicolumn{2}{c}{\textbf{Gemini-3.1-Pro}} & \multicolumn{2}{c}{\textbf{GPT-5.2}} \\
\cmidrule(lr){2-3}\cmidrule(lr){4-5}\cmidrule(lr){6-7}
\textbf{Input variant} & \textbf{Score} & $\Delta$ & \textbf{Score} & $\Delta$ & \textbf{Score} & $\Delta$ \\ \midrule
Unperturbed                & 7.20 $\pm$ 0.00 & --    & 7.83 $\pm$ 0.58 & --    & 7.70 $\pm$ 0.10 & --    \\ \midrule
JSON key order (3 shuffles)& 7.20 $\pm$ 0.00 & $+$0.00 & 7.92 $\pm$ 0.36 & $+$0.09 & 7.68 $\pm$ 0.22 & $-$0.02 \\
Orientation (5 transforms) & 7.20 $\pm$ 0.00 & $+$0.00 & 8.02 $\pm$ 0.26 & $+$0.19 & 7.81 $\pm$ 0.18 & $+$0.11 \\
\bottomrule
\end{tabular}
}
\end{table}
 
\begin{table}[htbp]
\centering
\caption{Robustness of the full tuning loop to input perturbations (Gemini-3.1-Pro-Preview, UMAP, healthy human kidney), with the perturbation reapplied at every iteration. Mean $\pm$ sd over 9 independent runs.}
\label{tab:robust_loop}
\small
\resizebox{\ifdim\width>\linewidth\linewidth\else\width\fi}{!}{
\begin{tabular}{@{}lccccc@{}}
\toprule
\textbf{Input variant} & \textbf{Impl. init.} & \textbf{Impl. best} & \textbf{Expl. best} & \textbf{Impl. final} & \textbf{Iters} \\ \midrule
Unperturbed            & 7.72 $\pm$ 0.26 & 8.81 $\pm$ 0.47 & 8.64 $\pm$ 0.16 & 8.47 $\pm$ 0.88 & 8.0 $\pm$ 3.3 \\ \midrule
JSON key order         & 7.73 $\pm$ 0.52 & 8.76 $\pm$ 0.34 & 8.67 $\pm$ 0.18 & 8.67 $\pm$ 0.33 & 6.2 $\pm$ 2.9 \\
Orientation (rot.\ 90) & 7.83 $\pm$ 0.25 & 8.72 $\pm$ 0.38 & 8.62 $\pm$ 0.19 & 8.32 $\pm$ 0.37 & 5.4 $\pm$ 3.0 \\
\bottomrule
\end{tabular}
}
\end{table}

\subsection{Further on LLM  Behaviors}

\textit{Gemini-3.1-Pro-Preview}  stops in 5.2 iterations on average and reaches the highest best implicit score in the majority of settings. In the representative trajectory shown, the peak coincides with a minimization of Stress-1 (0.323) and a high Spearman correlation (0.773), indicating good global structure recovery. Gemini is also the most parsimonious in its edits, changing 3.1 parameters per iteration on average against 5.2 for Claude and 5.5 for GPT-5.2.

\textit{GPT-5.2}  runs the longest search, 8.4 iterations on average, with visibly more volatile metric trajectories, most notably in Stress-1 and Spearman correlation. This is consistent across seeds and is the one behavioral difference besides Claude's scoring offset that exceeds the run-to-run spread.

\textit{Claude-Opus-4.5} stops earliest, after 2.1 iterations on average. Its scores are also systematically lower than those of the other two providers for the identical embedding, by 0.5 to 1.5 points across all 18 settings, so its early stopping reflects a stricter and more stable scoring behavior rather than a lower-quality result: its explicit composite scores are comparable to those of the other providers. Notably, despite assigning the highest explicit weight to the Silhouette Score (35\%), the  chosen hyperparameters result in a steady decline in both Trustworthiness and Silhouette scores, favoring a linear increase in Spearman Correlation instead.

\subsection{LLM-generated Score Weights}

\begin{lstlisting}[basicstyle=\ttfamily\scriptsize, breaklines=true, breakatwhitespace=true, frame=single, caption={Explicit scoring weights generated by LLMs.}, label={lst:model_weights}]
MODEL_METRIC_WEIGHTS = {
    "gpt-5.2": {
        "Trustworthiness":      0.30,
        "Silhouette Score":     0.30,
        "Spearman Correlation": 0.20,
        "Stress-1":             0.15,
        "LOF Median":           0.05,},
    "claude-opus-4-5": {
        "Trustworthiness":      0.20,
        "Silhouette Score":     0.35,
        "Spearman Correlation": 0.15,
        "Stress-1":             0.10,
        "LOF Median":           0.20,},
    "gemini-3.1-pro-preview": {
        "Trustworthiness":      0.25,
        "Silhouette Score":     0.25,
        "Spearman Correlation": 0.20,
        "Stress-1":             0.15,
        "LOF Median":           0.15,}}
\end{lstlisting}
 
\subsection{Implicit vs. Explicit Scoring Dynamics}
In Figure~\ref{fig:optimization_trajectories}, the dual-line plots (top row of each figure) highlight the positive correlation between mathematically rigid objective functions and holistic visual evaluation. 
Across all models, there is a visible ``disagreement gap'' (shaded in grey), particularly in the early iterations. For instance, Claude-Opus-4.5 initially assigns a harsh implicit score of 5.50, while the mathematical composite score rates the same embedding at 7.08. 
This divergence underscores the limitation of pure mathematical scoring: while the metrics may indicate a mathematically sound embedding, the LLM's implicit score penalizes visual artifacts (such as crowding or artificial cluster bridges) that the composite formula fails to capture.

\subsection{Trade-offs in Quantitative Metrics}

The metric subplots in Figure~\ref{fig:optimization_trajectories} reveal performance trade-offs inherent in dimensionality reduction. 
In almost all trajectories, improving global topology (reflected by decreasing Stress-1 and increasing Spearman correlation) comes at the direct expense of local structure (Trustworthiness and Silhouette Score). 
For example, in the Gemini trajectory, as Stress-1 reaches its absolute minimum at iteration 3 (0.323), the Silhouette Score simultaneously hits a severe low (0.013). 
This confirms that forcing the DR algorithm to perfectly unroll the global manifold often shatters tight local clusters, a trade-off the LLM agents have to  navigate dynamically throughout the pipeline.

 \subsection{Trajectory History and Memory}
\label{app:history}
The agent is trajectory-aware. For Claude and Gemini, the full history of previous iterations (parameters, scores and recommendations) is injected into the prompt at every iteration, with an instruction to reason about the observed effect of the previous change and not to re-propose values already tried. For GPT, the same information is carried server-side by chaining the previous response identifier.
Removing this history (Table~\ref{tab:input_ablation}, row ``w/o iteration history'') reduces the best implicit score from 9.00 to 8.00 and leaves the explicit best equal to the initial value, because the agent repeatedly proposes the same first-step changes. Iteration history is therefore the most important input component for tuning progress.

\section{Example of LLM Input Prompt}
\label{app:input}

\begin{figure}[htbp]
  \centering
  \includegraphics[width=0.85\linewidth,height=0.6\textheight,keepaspectratio]{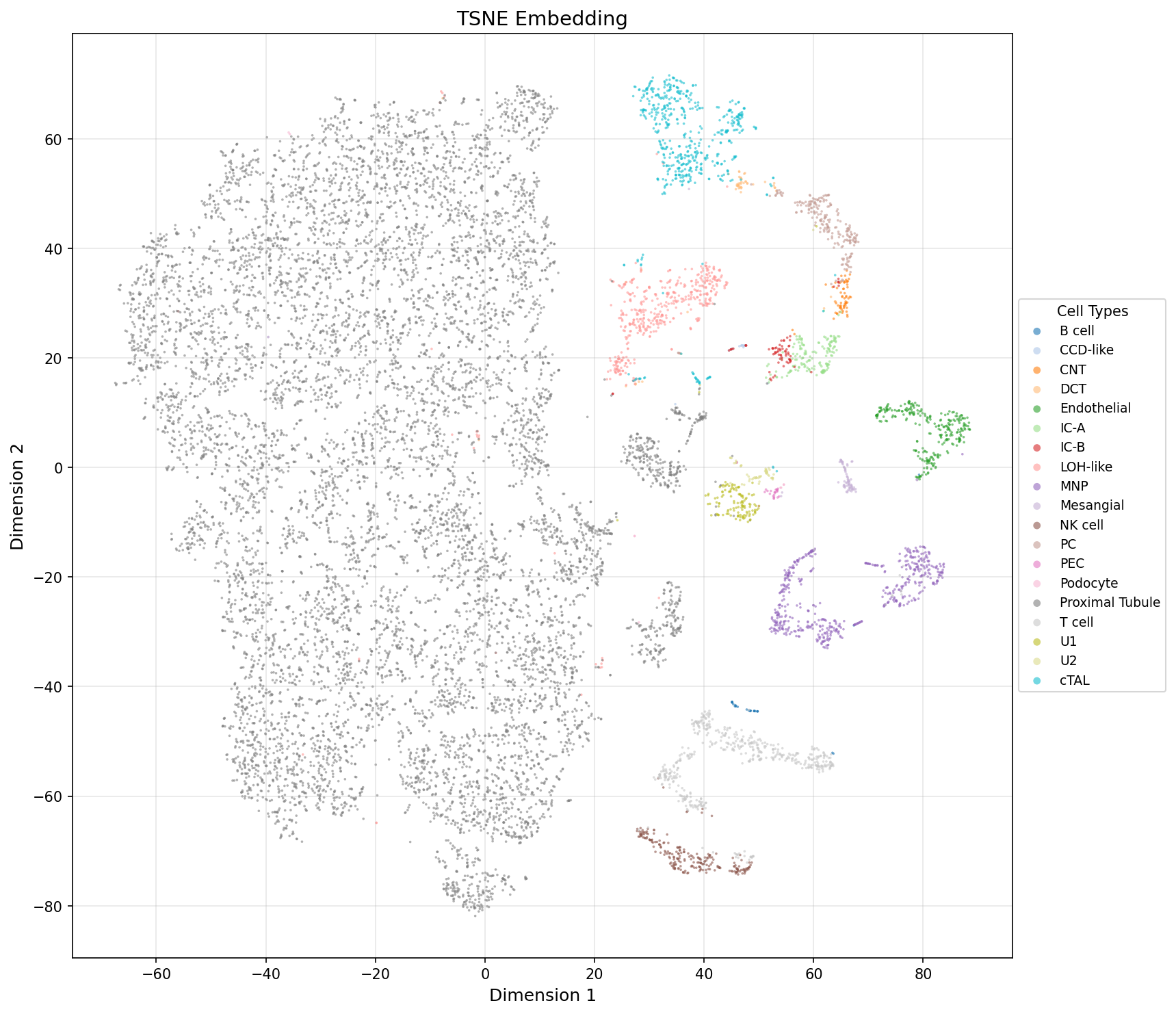}
  \caption{Embedding attached to the request at iteration 1 (\texttt{perplexity}=30, \texttt{learning\_rate}=200, \texttt{n\_iter}=1000, \texttt{n\_pcs}=20); 13,838 cells colored by \texttt{Cell\_Types\_Broad}.}
  \label{fig:ex_embedding}
\end{figure}

\begin{figure}[htbp]
  \centering
  \begin{subfigure}[b]{\columnwidth}
    \centering
    \includegraphics[width=\linewidth,height=0.36\textheight,keepaspectratio]{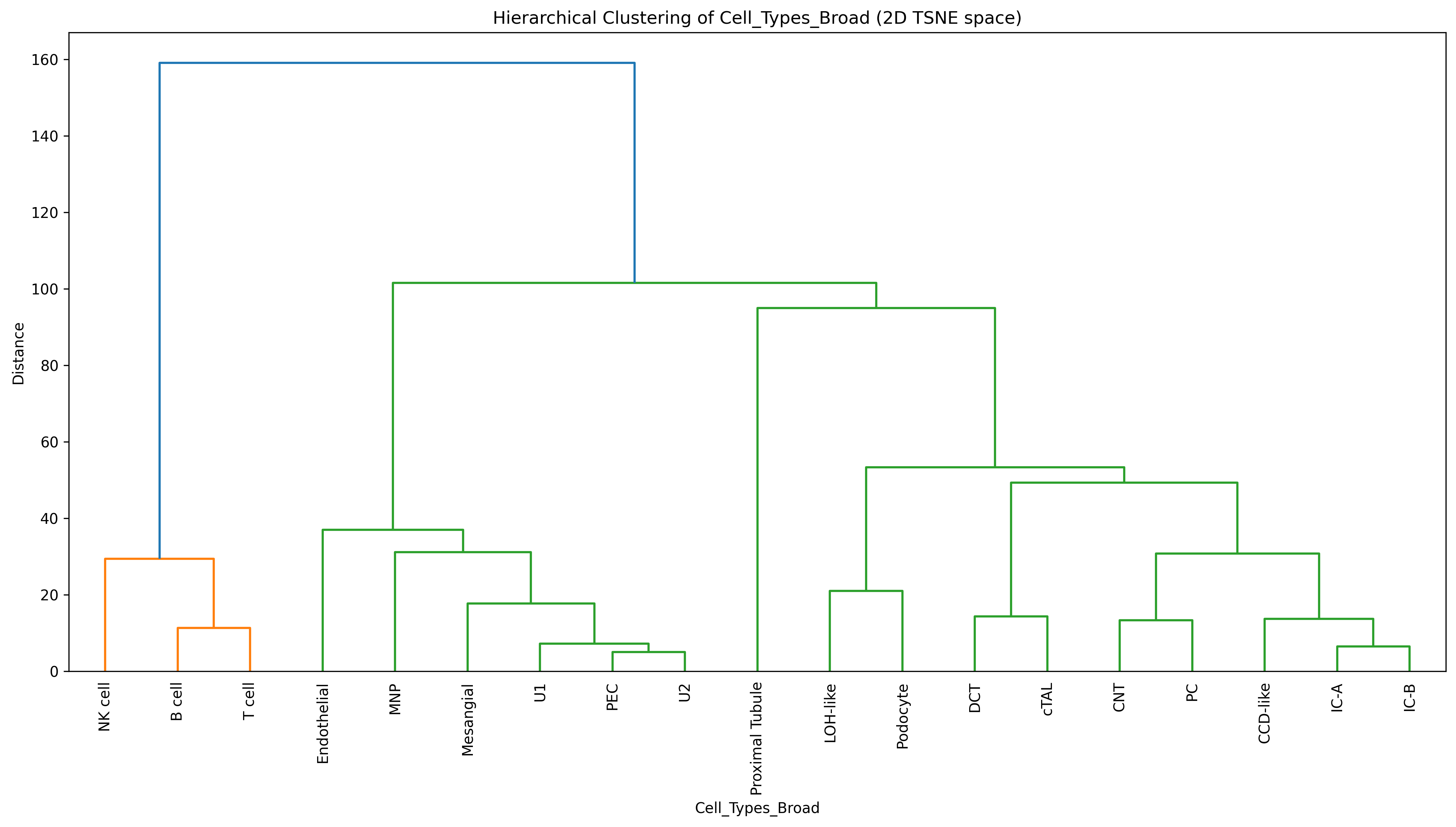}
    \caption{2D embedding}
    \label{fig:ex_dendro_2d}
  \end{subfigure}
  \vspace{4pt}
  \begin{subfigure}[b]{\columnwidth}
    \centering
    \includegraphics[width=\linewidth,height=0.36\textheight,keepaspectratio]{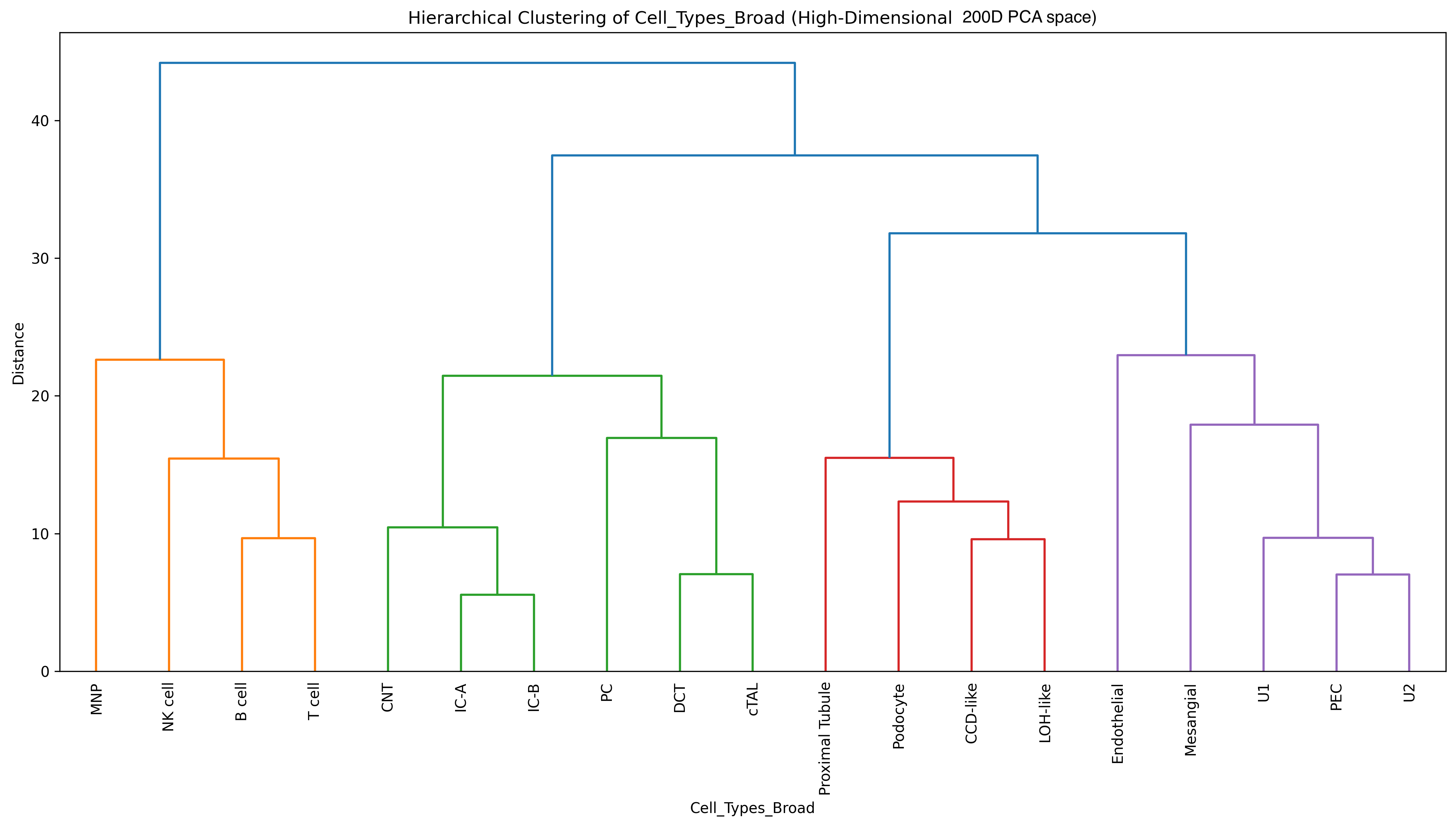}
    \caption{High-dimensional (200 PCs)}
    \label{fig:ex_dendro_hd}
  \end{subfigure}
  \caption{Dendrograms over cell-type centroids supplied to the agent at iteration 1, the graphical counterparts of the \texttt{hierarchy\_2d} and \texttt{hierarchy\_hd} blocks of Listing~\ref{lst:example_input}.}
  \label{fig:ex_dendro}
\end{figure}

We demonstrate an example of the complete input file for prompting the LLM.
The input has three parts: the instruction prompt (Listing~\ref{lst:example_prompt}), which is assembled at every iteration from the current parameter values, the label distribution, the metric block and the two dendrograms; the structured JSON payload (Listing~\ref{lst:example_input}), whose top-level blocks are the units removed one at a time in the input ablation of Table~\ref{tab:input_ablation}; and the rendered embedding (Figure~\ref{fig:ex_embedding}), attached to the request as an image, together with the two dendrograms (Figures~\ref{fig:ex_dendro_2d} and~\ref{fig:ex_dendro_hd}), which are supplied both as figures and in Newick form.
Leaf indices in the Newick strings refer to the alphabetically sorted \texttt{labels} array of the same block.
The catalogue of tunable parameters is supplied for all three DR methods regardless of which is active, so that a recommendation to switch method remains expressible.

\begin{lstlisting}[basicstyle=\ttfamily\scriptsize, breaklines=true, breakatwhitespace=true, frame=single, literate={↑}{{$\uparrow$}}1 {↓}{{$\downarrow$}}1 {→}{{$\rightarrow$}}1 {↔}{{$\leftrightarrow$}}1 {×}{{$\times$}}1 {≤}{{$\leq$}}1 {²}{{$^2$}}1, caption={Instruction prompt of iteration 1, as sent to the LLM.}, label={lst:example_prompt}]
# Human Kidney Cell Embedding Analysis
 
## INSTRUCTIONS
- Be concise and specific in your analysis
- Provide actionable recommendations with exact parameter values
- Focus on practical improvements that can be implemented immediately
- Explain the expected impact of each suggestion
- Prioritize recommendations by potential impact (high/medium/low)
 
## Dataset Overview
- **Label Column**: Cell_Types_Broad
- **Total Cell Types**: 19
- **Total Cells**: 13838
- **Embedding Method**: TSNE
 
## Method Parameters Used
  - method: tsne
  - n_pcs: 20
  - n_samples_used: 13838
  - random_state: 42
  - perplexity: 30.0
  - learning_rate: 200.0
  - n_iter: 1000
  - early_exaggeration: 12.0
  - tsne_metric: euclidean
  - tsne_init: pca
  - angle: 0.5
  - tsne_method: barnes_hut
  - n_iter_without_progress: 300
  - min_grad_norm: 1e-07
 
## Available Parameters for Recommendations
### UMAP Parameters (use these exact names in recommendations):
- **n_neighbors** (int): Local neighborhood size. ↑ = more global structure
- **min_dist** (float): Min distance between points. ↓ = tighter clusters
- **spread** (float): Scale of embedded points. ↑ = more spread
- **n_epochs** (int or None): Training iterations. None = auto
- **metric** (str): Distance metric for neighbors (euclidean, manhattan, cosine, correlation)
- **alpha** (float): Initial learning rate for UMAP optimization
- **gamma** (float): Repulsive force weight in cost function. ↑ = more repulsion
- **negative_sample_rate** (int): Negative samples per positive. ↑ = better separation
- **init_pos** (str): Initialization (spectral, paga, random)
- **n_pcs** (int): PCA components to use
 
### t-SNE Parameters (use these exact names in recommendations):
- **perplexity** (float): Balance local/global. ↑ = more global focus
- **learning_rate** (float): Optimization step size
- **n_iter** (int): Max iterations. ↑ = better convergence
- **early_exaggeration** (float): Cluster tightness in early phase. ↑ = tighter clusters
- **tsne_metric** (str): Distance metric (euclidean, manhattan, cosine, correlation)
- **tsne_init** (str): Initialization (pca, random)
- **angle** (float): Barnes-Hut trade-off. ↓ = more accurate but slower
- **tsne_method** (str): Computation (barnes_hut, exact). exact = more accurate but O(N²)
- **n_iter_without_progress** (int): Early stopping patience
- **min_grad_norm** (float): Convergence threshold
- **n_pcs** (int): PCA components to use
 
### PaCMAP Parameters (use these exact names in recommendations):
- **pacmap_n_neighbors** (int): Number of neighbors for local structure. ↑ = smoother layout, typical: 5-30
- **mn_ratio** (float): Mid-near pair ratio. Controls local vs global balance. ↑ = more local focus, typical: 0.1-2.0
- **fp_ratio** (float): Further pair ratio. Controls global structure preservation. ↑ = stronger global structure, typical: 0.5-5.0
- **pacmap_num_iters** (int): Total iterations across all 3 optimization phases, typical: 200-1500
- **pacmap_lr** (float): Learning rate for optimization. ↑ = faster but less stable, typical: 0.1-5.0
- **pacmap_distance** (str): Distance metric (euclidean, manhattan, angular, hamming, dot). Note: 'angular' is the cosine-based metric; do NOT use 'cosine'
- **pacmap_apply_pca** (bool): Whether to apply PCA preprocessing before PaCMAP (true/false)
- **n_pcs** (int): PCA components for preprocessing (used when pacmap_apply_pca is true)
 
## Cell Type Distribution
- Proximal Tubule: 10392 cells
- cTAL: 542 cells
- MNP: 528 cells
- LOH-like: 481 cells
- T cell: 478 cells
- Endothelial: 283 cells
- NK cell: 226 cells
- PC: 219 cells
- IC-A: 159 cells
- U1: 126 cells
- Mesangial: 76 cells
- IC-B: 74 cells
- CNT: 64 cells
- U2: 57 cells
- DCT: 53 cells
- B cell: 28 cells
- PEC: 25 cells
- CCD-like: 18 cells
- Podocyte: 9 cells
 
## Quantitative Metrics
{
  "Global Structure & HD↔2D Correlation": {
    "Mean 2D/HD Distance Ratio": "4.0426",
    "Spearman Correlation (Normalized Distances)": "0.5564",
    "Stress-1 (Normalized Distances)": "0.5856"
  },
  "Neighborhood Preservation": {
    "Trustworthiness (k=10)": "0.9952"
  },
  "Clustering & Label Agreement": {
    "Silhouette Score (True Labels)": "0.0557"
  },
  "Outlier Detection": {
    "LOF Median Score": "1.0127",
    "LOF Top 1% Score": "1.4183"
  },
  "Cluster Centroids (2D)": {
    "Centroid (Label B cell)": "(50.3311, -42.4053)",
    "Centroid (Label CCD-like)": "(44.6639, 20.5359)",
    "Centroid (Label CNT)": "(65.0610, 30.9315)",
    "Centroid (Label DCT)": "(43.4194, 44.9588)",
    "Centroid (Label Endothelial)": "(80.7375, 6.7651)",
    "Centroid (Label IC-A)": "(59.7583, 20.3971)",
    "Centroid (Label IC-B)": "(53.3678, 21.5515)",
    "Centroid (Label LOH-like)": "(30.3637, 26.5520)",
    "Centroid (Label MNP)": "(66.7975, -22.9626)",
    "Centroid (Label Mesangial)": "(63.3512, -0.7589)",
    "Centroid (Label NK cell)": "(37.4570, -70.7618)",
    "Centroid (Label PC)": "(61.9978, 43.9768)",
    "Centroid (Label PEC)": "(52.1183, -4.2928)",
    "Centroid (Label Podocyte)": "(25.3749, 6.1138)",
    "Centroid (Label Proximal Tubule)": "(-16.5477, -1.3214)",
    "Centroid (Label T cell)": "(46.4841, -53.1081)",
    "Centroid (Label U1)": "(45.7856, -5.9486)",
    "Centroid (Label U2)": "(49.4529, -0.0447)",
    "Centroid (Label cTAL)": "(37.0862, 57.8384)"
  },
  "Cluster Centroid Separation": {
    "Mean Centroid Distance": "48.8124",
    "Std Centroid Distance": "26.6236"
  }
}
 
## Hierarchical Structure Analysis
 
### 2D Embedding Dendrogram (Based on TSNE coordinates)
**Newick Format:**
((10,(0,15)),((4,(8,(9,(16,(12,17))))),
(14,((7,13),((3,18),((2,11),(1,(5,6))))))));
 
**Summary Statistics:**
{
  "n_leaves": 19,
  "max_height": 159.04005535023884,
  "median_height": 25.229113017423327,
  "clusters_at_height≤5": 5,
  "labels": ["B cell", "CCD-like", "CNT", "DCT", "Endothelial", "IC-A", "IC-B", "LOH-like", "MNP", "Mesangial", "NK cell", "PC", "PEC", "Podocyte", "Proximal Tubule", "T cell", "U1", "U2", "cTAL"],
  "label_column": "Cell_Types_Broad",
  "dimension": "2D"
}
 
### High-Dimensional Dendrogram (Based on PCA space)
**Newick Format:**
((8,(10,(0,15))),(((2,(5,6)),(11,(3,18))),
((14,(13,(1,7))),(4,(9,(16,(12,17)))))));
 
**Summary Statistics:**
{
  "n_leaves": 19,
  "max_height": 44.17237406864259,
  "median_height": 15.483944064435718,
  "clusters_at_height≤5": 5,
  "labels": ["B cell", "CCD-like", "CNT", "DCT", "Endothelial", "IC-A", "IC-B", "LOH-like", "MNP", "Mesangial", "NK cell", "PC", "PEC", "Podocyte", "Proximal Tubule", "T cell", "U1", "U2", "cTAL"],
  "label_column": "Cell_Types_Broad",
  "dimension": "20D (PCA)"
}
 
## Explicit Score Calculation
 
You MUST compute an **explicit_score** (0-10) using the following formula.
Extract each raw metric from the Quantitative Metrics section above, normalise, apply weights, and sum.
 
| Metric | JSON Path | Normalisation | Weight |
|--------|-----------|---------------|--------|
| Trustworthiness (k=10) | Neighborhood Preservation → Trustworthiness (k=10) | clip to [0, 1] | 0.25 |
| Silhouette Score (True Labels) | Clustering & Label Agreement → Silhouette Score (True Labels) | (value + 1) / 2, clip [0,1] | 0.25 |
| Spearman Correlation | Global Structure & HD↔2D Correlation → Spearman Correlation (Normalized Distances) | clip to [0, 1] | 0.2 |
| Stress-1 | Global Structure & HD↔2D Correlation → Stress-1 (Normalized Distances) | 1.0 - clip(value, 0, 1) (invert: lower stress = higher score) | 0.15 |
| LOF Median Score | Outlier Detection → LOF Median Score | max(0, 1 - (value - 1)) (closer to 1.0 = better) | 0.15 |
 
**Formula:** explicit_score = (sum of normalised_metric_i × weight_i) / (sum of weights for available metrics) × 10
 
Compute this score precisely from the numeric metrics provided. Report it in the `explicit_score` field.
 
## Analysis Tasks
 
Please analyze this single-cell RNA-seq embedding based on metrics, dendrograms, the visual plot.
 
**Your analysis should:**
1. Compute the explicit_score using the formula above (MANDATORY)
2. Assess embedding quality using the quantitative metrics provided
3. Compare 2D vs high-dimensional dendrogram structures to evaluate preservation
4. Evaluate visual cluster separation, compactness, and artifacts
5. Provide specific, actionable parameter recommendations with exact values
6. Explain expected impact for each recommendation (e.g., "Increasing n_neighbors from X to Y will improve global structure preservation")
7. Identify any biological insights or concerns about cell type relationships
8. Keep explanations concise and focused on actionable insights
 
**Format requirements:**
- Use concrete numbers and specific values in recommendations
- Prioritize suggestions by impact (high = major improvement, medium = moderate, low = minor refinement)
- Be direct and avoid vague statements
- Focus on what can be done, not just what's wrong
- The explicit_score MUST be computed from the metrics, not estimated
\end{lstlisting}
 
\begin{lstlisting}[basicstyle=\ttfamily\scriptsize, breaklines=true, breakatwhitespace=true, frame=single, literate={↔}{{$\leftrightarrow$}}1 {≤}{{$\leq$}}1, caption={Structured JSON input of iteration 1.}, label={lst:example_input}]
{
  "metrics": {
    "Global Structure & HD↔2D Correlation": {
      "Mean 2D/HD Distance Ratio": "4.0426",
      "Spearman Correlation (Normalized Distances)": "0.5564",
      "Stress-1 (Normalized Distances)": "0.5856"
    },
    "Neighborhood Preservation": {
      "Trustworthiness (k=10)": "0.9952"
    },
    "Clustering & Label Agreement": {
      "Silhouette Score (True Labels)": "0.0557"
    },
    "Outlier Detection": {
      "LOF Median Score": "1.0127",
      "LOF Top 1% Score": "1.4183"
    },
    "Cluster Centroids (2D)": {
      "Centroid (Label B cell)": "(50.3311, -42.4053)",
      "Centroid (Label CCD-like)": "(44.6639, 20.5359)",
      "Centroid (Label CNT)": "(65.0610, 30.9315)",
      "Centroid (Label DCT)": "(43.4194, 44.9588)",
      "Centroid (Label Endothelial)": "(80.7375, 6.7651)",
      "Centroid (Label IC-A)": "(59.7583, 20.3971)",
      "Centroid (Label IC-B)": "(53.3678, 21.5515)",
      "Centroid (Label LOH-like)": "(30.3637, 26.5520)",
      "Centroid (Label MNP)": "(66.7975, -22.9626)",
      "Centroid (Label Mesangial)": "(63.3512, -0.7589)",
      "Centroid (Label NK cell)": "(37.4570, -70.7618)",
      "Centroid (Label PC)": "(61.9978, 43.9768)",
      "Centroid (Label PEC)": "(52.1183, -4.2928)",
      "Centroid (Label Podocyte)": "(25.3749, 6.1138)",
      "Centroid (Label Proximal Tubule)": "(-16.5477, -1.3214)",
      "Centroid (Label T cell)": "(46.4841, -53.1081)",
      "Centroid (Label U1)": "(45.7856, -5.9486)",
      "Centroid (Label U2)": "(49.4529, -0.0447)",
      "Centroid (Label cTAL)": "(37.0862, 57.8384)"
    },
    "Cluster Centroid Separation": {
      "Mean Centroid Distance": "48.8124",
      "Std Centroid Distance": "26.6236"
    }
  },
  "label_summary": {
    "Proximal Tubule": 10392, "NK cell": 226, "LOH-like": 481,
    "Endothelial": 283, "T cell": 478, "cTAL": 542, "IC-A": 159,
    "U1": 126, "DCT": 53, "MNP": 528, "Mesangial": 76, "CNT": 64,
    "PC": 219, "IC-B": 74, "U2": 57, "CCD-like": 18, "PEC": 25,
    "B cell": 28, "Podocyte": 9
  },
  "label_column": "Cell_Types_Broad",
  "hierarchy_2d": {
    "newick": "((10,(0,15)),((4,(8,(9,(16,(12,17)))))
    ,(14,((7,13),((3,18),((2,11),(1,(5,6))))))));",
    "n_leaves": 19,
    "max_height": 159.04005535023884,
    "median_height": 25.229113017423327,
    "clusters_at_height≤5": 5,
    "labels": ["B cell", "CCD-like", "CNT", "DCT", "Endothelial", "IC-A", "IC-B", "LOH-like", "MNP", "Mesangial", "NK cell", "PC", "PEC", "Podocyte", "Proximal Tubule", "T cell", "U1", "U2", "cTAL"],
    "label_column": "Cell_Types_Broad",
    "dimension": "2D"
  },
  "hierarchy_hd": {
    "newick": "((8,(10,(0,15))),(((2,(5,6)),
    (11,(3,18))),((14,(13,(1,7))),
    (4,(9,(16,(12,17)))))));",
    "n_leaves": 19,
    "max_height": 44.17237406864259,
    "median_height": 15.483944064435718,
    "clusters_at_height≤5": 5,
    "labels": ["B cell", "CCD-like", "CNT", "DCT", "Endothelial", "IC-A", "IC-B", "LOH-like", "MNP", "Mesangial", "NK cell", "PC", "PEC", "Podocyte", "Proximal Tubule", "T cell", "U1", "U2", "cTAL"],
    "label_column": "Cell_Types_Broad",
    "dimension": "200D (PCA)"
  },
  "method": "tsne",
  "parameters": {
    "method": "tsne",
    "n_pcs": 20,
    "n_samples_used": 13838,
    "random_state": 42,
    "perplexity": 30.0,
    "learning_rate": 200.0,
    "n_iter": 1000,
    "early_exaggeration": 12.0,
    "tsne_metric": "euclidean",
    "tsne_init": "pca",
    "angle": 0.5,
    "tsne_method": "barnes_hut",
    "n_iter_without_progress": 300,
    "min_grad_norm": 1e-07
  }
}
\end{lstlisting}
 
\section{Example of LLM Output Response}
\label{app:output}

We demonstrate an example of the complete output response file returned by the LLM, including its diagnosis and analysis of the initial hyperparameter configuration, and recommended hyperparameter changes along with their rationale across iterations, as well as the explanation of the visualization output.

\begin{figure}[htbp]
  \centering
  \includegraphics[width=0.85\linewidth,height=0.6\textheight,keepaspectratio]{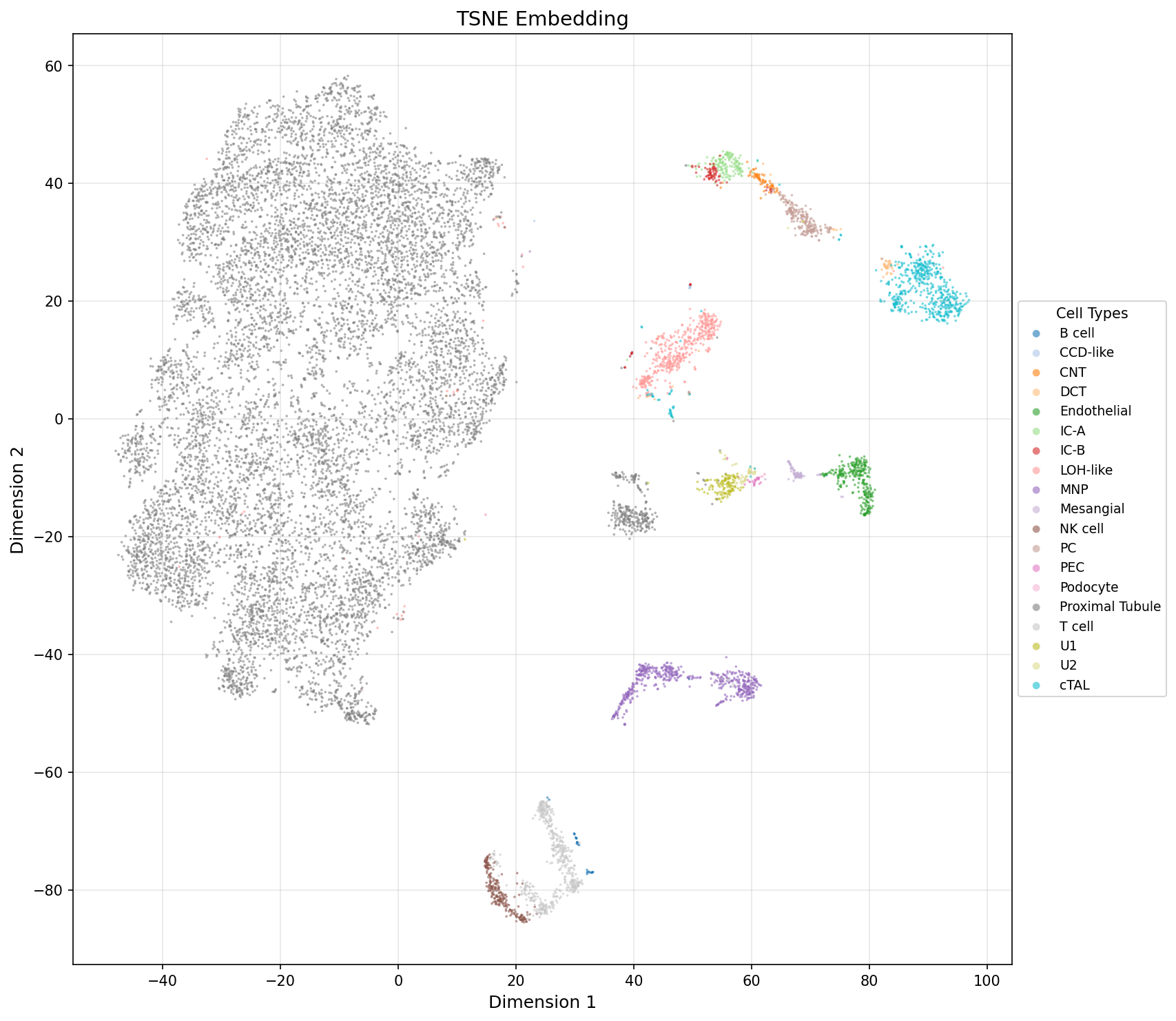}
  \caption{Embedding obtained at iteration 2 after applying the recommendations of Listing~\ref{lst:example_output} (\texttt{perplexity}=100, \texttt{learning\_rate}=800, \texttt{n\_iter}=2000, \texttt{early\_exaggeration}=24, \texttt{n\_pcs}=20); implicit score 7.5.}
  \label{fig:ex_embedding_it2}
\end{figure}

\begin{figure}[htbp]
  \centering
  \begin{subfigure}[b]{\columnwidth}
    \centering
    \includegraphics[width=\linewidth,height=0.36\textheight,keepaspectratio]{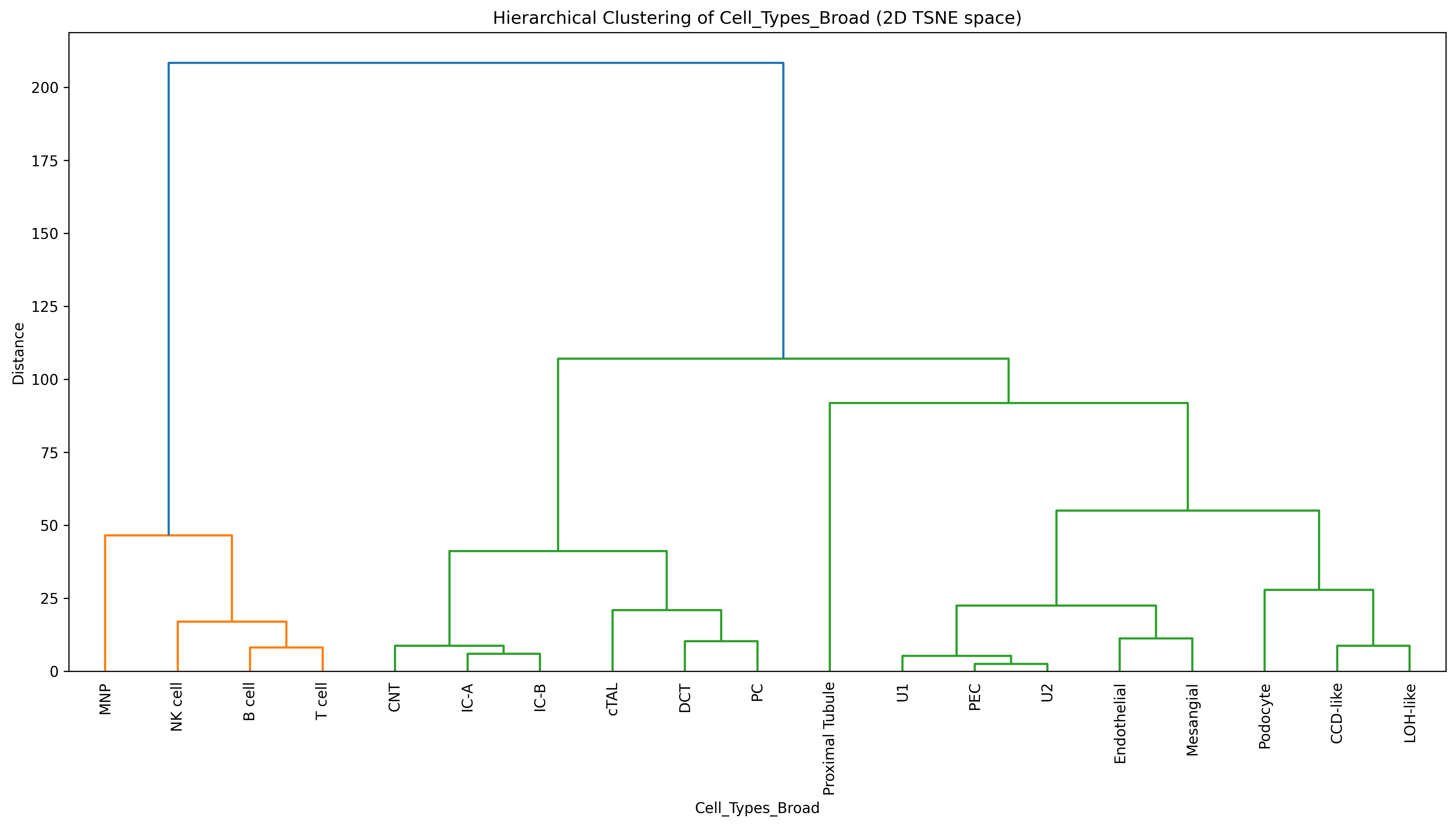}
    \caption{2D embedding}
    \label{fig:ex_dendro_2d_it2}
  \end{subfigure}
  \vspace{4pt}
  \begin{subfigure}[b]{\columnwidth}
    \centering
    \includegraphics[width=\linewidth,height=0.36\textheight,keepaspectratio]{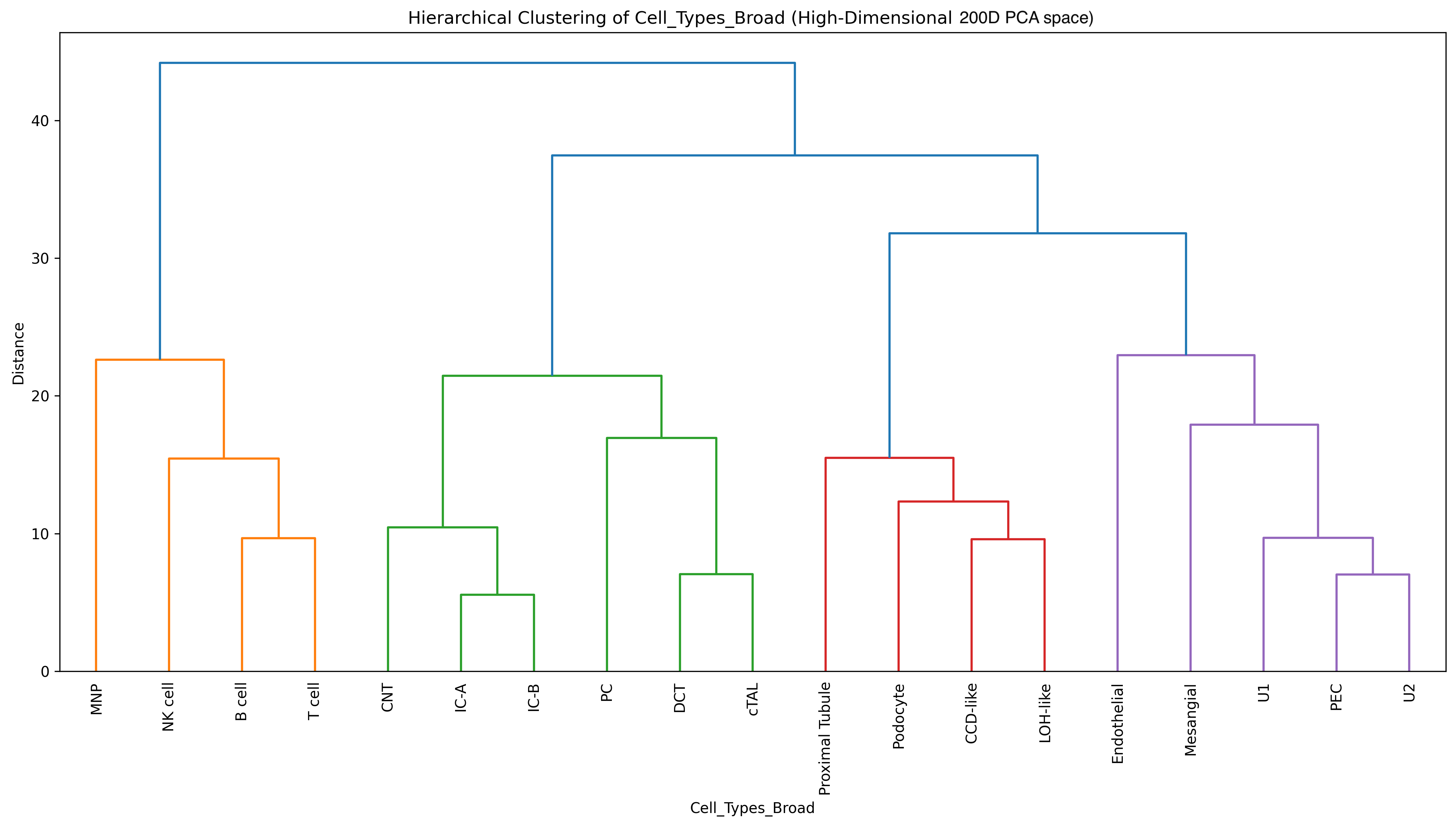}
    \caption{High-dimensional (200 PCs)}
    \label{fig:ex_dendro_hd_it2}
  \end{subfigure}
  \caption{Dendrograms over cell-type centroids at iteration 2, supplied to the agent together with Figure~\ref{fig:ex_embedding_it2} as the input of the next step.}
  \label{fig:ex_dendro_it2}
\end{figure}

Listing~\ref{lst:example_output} is the agent's response to the input above, returned as a single JSON document against the schema of Listing~\ref{lst:output_response}.
It contains the implicit score, six analysis blocks (\texttt{score\_rationale}, \texttt{overall\_assessment}, \texttt{dendrogram\_comparison}, \texttt{visual\_inspection}, \texttt{cell\_type\_relationships}, \texttt{outlier\_analysis}), the parameter recommendations with rationale and expected impact, and the proposed follow-up metrics.
Only the \texttt{recommendations} array is consumed by the pipeline to build the next parameter set; every other field is explanatory and is what the external evaluation assesses statement by statement.
Applying the four recommendations gives iteration 2 (\texttt{perplexity} $30 \rightarrow 100$, \texttt{early\_exaggeration} $12 \rightarrow 24$, \texttt{learning\_rate} $200 \rightarrow 800$, \texttt{n\_iter} $1000 \rightarrow 2000$); the resulting embedding and its dendrograms are shown in Figures~\ref{fig:ex_embedding_it2} and~\ref{fig:ex_dendro_it2}, and the input document is regenerated from them in the same form, with the history of previous iterations added as described in Appendix~\ref{app:history}.
The remaining iterations of this trajectory follow the same format and are omitted for space.

\begin{lstlisting}[basicstyle=\ttfamily\scriptsize, breaklines=true, breakatwhitespace=true, frame=single, caption={Structured output of iteration 1.}, label={lst:example_output}]
{
  "quality_score": 6,
  "score_rationale": "The embedding achieves excellent local neighborhood preservation (Trustworthiness ~0.99) but suffers from poor cluster separation (Silhouette ~0.05) and a lack of distinct global structure. The visualization is dominated by a single massive, amorphous cluster (Proximal Tubule) with insufficient separation for minority cell types.",
  "overall_assessment": {
    "summary": "The t-SNE embedding is heavily influenced by the extreme class imbalance (75% Proximal Tubule). While local relationships are trusted, the low perplexity and standard exaggeration parameters have failed to separate distinct biological populations effectively, resulting in a 'hairball' structure for the majority class and fragmentation of minority classes.",
    "key_strengths": [
      "Excellent preservation of local neighborhoods (Trustworthiness > 0.99)",
      "Broad biological lineages (Immune vs Nephron) are roughly spatially distinct",
      "Low outlier scores suggest data quality is acceptable"
    ],
    "key_weaknesses": [
      "Extremely low silhouette score indicates significant cluster overlap",
      "Majority class (Proximal Tubule) lacks internal structure definition",
      "Global structure metric (Spearman ~0.55) indicates distortion of long-range distances",
      "Minority clusters are somewhat scattered and not clearly defined"
    ],
    "metric_analysis": {
      "trustworthiness": "0.9952 (Excellent) - Local neighbors are highly consistent between HD and 2D space.",
      "stress": "0.5856 (Moderate) - Indicates some difficulty compressing the high-dimensional variance into 2D.",
      "silhouette": "0.0557 (Poor) - Clusters are not distinct; high overlap or fragmentation relative to labels.",
      "other_metrics": "Mean Centroid Distance (48.8) is relatively low given the scale, confirming tight packing/overlap."
    }
  },
  "dendrogram_comparison": {
    "agreement_level": "high",
    "key_similarities": [
      "Immune cells (B cell, T cell, NK cell) cluster together in both 2D and HD hierarchies.",
      "Nephron lineage components (Proximal Tubule, Podocyte, LOH-like) maintain a consistent branch grouping.",
      "MNP (Mononuclear Phagocytes) associate with the immune group in both."
    ],
    "key_differences": [
      "The placement of Endothelial cells varies slightly in branching depth relative to the stromal compartment.",
      "The internal hierarchy of the smaller 'U' (Unknown) clusters shifts between representations."
    ],
    "structure_preservation": "The embedding faithfully preserves the high-level biological hierarchy (Immune vs Structural vs Nephron) despite the visual crowding."
  },
  "visual_inspection": {
    "cluster_separation": "poor",
    "cluster_compactness": "moderate",
    "artifacts": [
      "Large, amorphous 'hairball' structure for Proximal Tubule",
      "Fragmentation of the cTAL and DCT clusters"
    ],
    "distribution_uniformity": "The plot is highly skewed, with dense packing on the left and sparse clusters on the right.",
    "notable_patterns": [
      "Immune cells form the most distinct separated islands on the right.",
      "The transition from Proximal Tubule to other nephron segments is not visually continuous, suggesting potential gaps in the trajectory or embedding artifacts."
    ]
  },
  "cell_type_relationships": {
    "biological_relevance": "Broadly consistent with kidney biology, separating immune infiltrates from parenchymal cells.",
    "expected_groupings": [
      "T cells, B cells, and NK cells clustering nearby (observed)",
      "Podocytes and Glomerular cells clustering near Proximal Tubule (observed)"
    ],
    "unexpected_groupings": [
      "cTAL and DCT appear somewhat fragmented rather than forming a clean distal nephron trajectory.",
      "Mesangial cells are somewhat isolated from other structural cells."
    ],
    "spatial_organization": "The embedding separates the dominant epithelial compartment (left) from the interstitial/immune compartment (right), but lacks fine-grained resolution within the epithelial block."
  },
  "outlier_analysis": {
    "lof_assessment": "Stable. Median LOF near 1.0 suggests most cells are in well-supported density regions.",
    "concerning_patterns": [
      "No severe outliers detected.",
      "Some sparse points at the periphery of the Proximal Tubule cluster likely represent transitional states or noise."
    ],
    "data_quality_issues": [
      "Extreme class imbalance is the primary analytical challenge, not necessarily data quality."
    ]
  },
  "recommendations": [
    {
      "parameter": "perplexity",
      "current_value": "30.0",
      "suggested_value": "100",
      "rationale": "With ~13k cells and a massive continuous cluster, higher perplexity is needed to model global structure and density variations accurately.",
      "expected_impact": "Better global organization and continuity of the large Proximal Tubule cluster; reduced fragmentation of smaller clusters.",
      "priority": "high"
    },
    {
      "parameter": "early_exaggeration",
      "current_value": "12.0",
      "suggested_value": "24.0",
      "rationale": "Standard exaggeration is insufficient to create white space between clusters in such a dense dataset.",
      "expected_impact": "Significantly improved cluster separation and higher Silhouette score; clearer distinction between subtypes.",
      "priority": "high"
    },
    {
      "parameter": "learning_rate",
      "current_value": "200.0",
      "suggested_value": "800",
      "rationale": "For N > 10,000, the default learning rate (200) is often too low, leading to suboptimal convergence/clumping.",
      "expected_impact": "More stable convergence and potentially better expansion of the dense central cluster.",
      "priority": "medium"
    },
    {
      "parameter": "n_iter",
      "current_value": "1000",
      "suggested_value": "2000",
      "rationale": "Given the complex topology and recommended higher perplexity, more iterations ensure the optimization reaches a stable minimum.",
      "expected_impact": "Refined cluster shapes and more reliable final positioning.",
      "priority": "medium"
    }
  ],
  "follow_up_metrics": [
    {
      "metric_name": "Silhouette Score",
      "description": "Measure of how similar an object is to its own cluster compared to other clusters.",
      "expected_benefit": "Should increase towards >0.1 or 0.2 with higher early_exaggeration."
    },
    {
      "metric_name": "Spearman Correlation (HD-2D)",
      "description": "Correlation of pairwise distances.",
      "expected_benefit": "Should increase with higher perplexity, indicating better global structure preservation."
    }
  ]
}
\end{lstlisting}

\FloatBarrier
\vskip 0.2in
\bibliography{references}

@article{vandermaaten08a,
  author  = {van der Maaten, Laurens and Hinton, Geoffrey},
  title   = {Visualizing Data using t-{SNE}},
  journal = {Journal of Machine Learning Research},
  volume  = {9},
  number  = {86},
  pages   = {2579--2605},
  year    = {2008}
}

@article{mcinnes2020umap,
  author  = {McInnes, Leland and Healy, John and Melville, James},
  title   = {{UMAP}: Uniform Manifold Approximation and Projection for Dimension Reduction},
  journal = {arXiv preprint arXiv:1802.03426},
  year    = {2020}
}

@article{arbelaitz2013extensive,
  author  = {Arbelaitz, Olatz and Gurrutxaga, Ibai and Muguerza, Javier and P{\'e}rez, Jes{\'u}s M. and Perona, I{\~n}igo},
  title   = {An Extensive Comparative Study of Cluster Validity Indices},
  journal = {Pattern Recognition},
  volume  = {46},
  number  = {1},
  pages   = {243--256},
  year    = {2013}
}

@article{chari2023stop,
  author  = {Chari, Tara and Pachter, Lior},
  title   = {The Specious Art of Single-Cell Genomics},
  journal = {PLoS Computational Biology},
  volume  = {19},
  number  = {10},
  pages   = {e1011288},
  year    = {2023}
}

@article{wang2021understanding,
  author  = {Wang, Yingfan and Huang, Haiyang and Rudin, Cynthia and Shaposhnik, Yaron},
  title   = {Understanding How Dimension Reduction Tools Work: An Empirical Approach to Deciphering t-{SNE}, {UMAP}, {TriMAP}, and {PaCMAP} for Data Visualization},
  journal = {Journal of Machine Learning Research},
  volume  = {22},
  number  = {1},
  pages   = {9129--9201},
  year    = {2021}
}

@article{wattenberg2016how,
  author  = {Wattenberg, Martin and Vi{\'e}gas, Fernanda and Johnson, Ian},
  title   = {How to Use t-{SNE} Effectively},
  journal = {Distill},
  year    = {2016}
}

@article{venna2006local,
  author  = {Venna, Jarkko and Kaski, Samuel},
  title   = {Local Multidimensional Scaling},
  journal = {Neural Networks},
  volume  = {19},
  number  = {8},
  pages   = {1189--1199},
  year    = {2006}
}

@article{bergstra12a,
  author  = {Bergstra, James and Bengio, Yoshua},
  title   = {Random Search for Hyper-Parameter Optimization},
  journal = {Journal of Machine Learning Research},
  volume  = {13},
  number  = {10},
  pages   = {281--305},
  year    = {2012}
}

@inproceedings{snoek2012practical,
  author    = {Snoek, Jasper and Larochelle, Hugo and Adams, Ryan P.},
  title     = {Practical {B}ayesian Optimization of Machine Learning Algorithms},
  booktitle = {Advances in Neural Information Processing Systems},
  volume    = {25},
  pages     = {2951--2959},
  year      = {2012}
}

@article{wang2023survey,
  author  = {Wang, Lei and others},
  title   = {A Survey on Large Language Model Based Autonomous Agents},
  journal = {Frontiers of Computer Science},
  volume  = {18},
  number  = {6},
  pages   = {1--26},
  year    = {2024}
}

@inproceedings{yang2024matplotagent,
  author    = {Yang, Zhiyu and others},
  title     = {{MatPlotAgent}: Method and Evaluation for {LLM}-Based Agentic Scientific Data Visualization},
  booktitle = {Findings of the Association for Computational Linguistics: ACL 2024},
  pages     = {11789--11804},
  address   = {Bangkok, Thailand},
  year      = {2024}
}

@article{chen2025coda,
  author  = {Chen, Z. and others},
  title   = {Coda: Agentic Systems for Collaborative Data Visualization},
  journal = {arXiv preprint arXiv:2510.03194},
  year    = {2025}
}

@inproceedings{liu2025agenthpo,
  author    = {Liu, Siyi and Gao, Chen and Li, Yong},
  title     = {{AgentHPO}: Large Language Model Agent for Hyper-Parameter Optimization},
  booktitle = {Proceedings of the Conference on Parsimony and Learning},
  series    = {Proceedings of Machine Learning Research},
  volume    = {280},
  pages     = {1146--1169},
  year      = {2025}
}

@article{sokal1958,
  author  = {Sokal, Robert R. and Michener, Charles D.},
  title   = {A Statistical Method for Evaluating Systematic Relationships},
  journal = {University of Kansas Science Bulletin},
  volume  = {38},
  pages   = {1409--1438},
  year    = {1958}
}

@article{rousseeuw1987,
  author  = {Rousseeuw, Peter J.},
  title   = {Silhouettes: A Graphical Aid to the Interpretation and Validation of Cluster Analysis},
  journal = {Journal of Computational and Applied Mathematics},
  volume  = {20},
  pages   = {53--65},
  year    = {1987}
}

@article{breunig2000lof,
  author  = {Breunig, Markus M. and Kriegel, Hans-Peter and Ng, Raymond T. and Sander, J{\"o}rg},
  title   = {{LOF}: Identifying Density-Based Local Outliers},
  journal = {SIGMOD Record},
  volume  = {29},
  number  = {2},
  pages   = {93--104},
  year    = {2000}
}

@article{spearman1904,
  author  = {Spearman, Charles},
  title   = {The Proof and Measurement of Association between Two Things},
  journal = {The American Journal of Psychology},
  volume  = {15},
  number  = {1},
  pages   = {72--101},
  year    = {1904}
}

@article{McEvoy2022,
  author  = {McEvoy, Caitriona M. and others},
  title   = {Single-Cell Profiling of Healthy Human Kidney Reveals Features of Sex-Based Transcriptional Programs and Tissue-Specific Immunity},
  journal = {Nature Communications},
  volume  = {13},
  number  = {1},
  pages   = {7634},
  year    = {2022}
}

@article{Stewart2019,
  author  = {Stewart, Benjamin J. and others},
  title   = {Spatiotemporal Immune Zonation of the Human Kidney},
  journal = {Science},
  volume  = {365},
  number  = {6459},
  pages   = {1461--1466},
  year    = {2019}
}

@article{speir2021ucsc,
  author  = {Speir, Matthew L. and others},
  title   = {{UCSC} Cell Browser: Visualize Your Single-Cell Data},
  journal = {Bioinformatics},
  volume  = {37},
  number  = {23},
  pages   = {4578--4580},
  year    = {2021}
}

\end{document}